\documentstyle[12pt,epsfig,subfigure,glas,ssi]{article}
\newlength{\dinwidth}                                                          
\newlength{\dinmargin}                                                         
\input epsf
\epsfverbosetrue
\begin{document}
\newcommand {\pom}  {I\hspace{-0.2em}P}
\newcommand {\regge}  {I\hspace{-0.2em}R}
\newcommand {\xpom} {\mbox{$x_{_{\pom}}$}}
\newcommand {\xipom} {\mbox{$x_{_{i/\pom}}$}}
\newcommand {\Spom} {\mbox{$\Sigma_{_{\pom}}$}}
\newcommand {\fpom} {\mbox{$F_2^{\pom}$}}
\newcommand {\alphapom} {\mbox{$\alpha_{_{\pom}}$}}
\newcommand {\alphappom} {\mbox{$\alpha'$}}
\newcommand {\fregge} {\mbox{$F_2^{\regge}$}}
\newcommand {\cregge} {\mbox{$C_{_{\regge}}$}}
\newcommand {\alpharegge} {\mbox{$\alpha_{_{\regge}}$}}
\newcommand {\alphapregge} {\mbox{$\alpha'_{_{\regge}}$}}
\newcommand{\sleq} {\raisebox{-.6ex}{${\textstyle\stackrel{<}{\sim}}$}}
\newcommand{\sgeq} {\raisebox{-.6ex}{${\textstyle\stackrel{>}{\sim}}$}}
\begin{titlepage}{GLAS-PPE/97--13}{19$^{\underline{\rm{th}}}$ December 1997}
\title{Diffraction: QCD Effects in Colour Singlet Exchange}
\author{Anthony T. Doyle\thanks{Supported by PPARC and DESY.
}\\
Department of Physics and Astronomy,\\
University of Glasgow.}
\begin{abstract}
Measurements of diffractive phenomena observed at HERA and the Tevatron 
are reviewed. A short introduction to the theoretical background is 
presented where colour singlet exchange reactions are discussed 
and the diffractive contribution and its interpretation
via pomeron exchange outlined.
The review focuses on the current experimental directions at HERA and 
discusses exclusive production of vector mesons, the dissociation of 
real photons and the deep inelastic structure of diffraction.
Complementary 
information obtained from hadronic final states and jet structures
is also discussed.
The experimental signatures for diffractive jet and $W^\pm$ production 
observed at the Tevatron are described and the rates compared with those
from the HERA experiments.

\vspace{0.5cm}
\centerline{\em Talk presented at the 25th SLAC Summer Institute,}
\centerline{\em SLAC, August 1997.}
\end{abstract}
\newpage
\end{titlepage}

\section{Introduction}
{\it{Prologue:}}
In 1992, H. Fritzsch summarised the status of 
``QCD 20 Years On"~\cite{fritzsch} 
with the words
``A very large amount of data in strong-interaction physics is described by
the pomeron singularity... 
The fact that the physics of the pomeron is very simple needs to be 
explained in equally simple terms in QCD...
Using HERA the experimentalists will be able to study the region of very
low $x$ in deep inelastic scattering... Furthermore, both in
$p\bar{p}$-scattering at high energies and at HERA the structure functions of
the pomeron can be studied in more detail."
Now, 25 years on from the first developments of QCD, 
at the 25th SLAC Summer Institute we can discuss
a series of diffractive measurements that have been made in the 
intervening period at HERA and the Tevatron. 

\begin{figure}[htb]
\epsfxsize=8.cm
\centering
\leavevmode
\epsfbox[110 270 480 570]{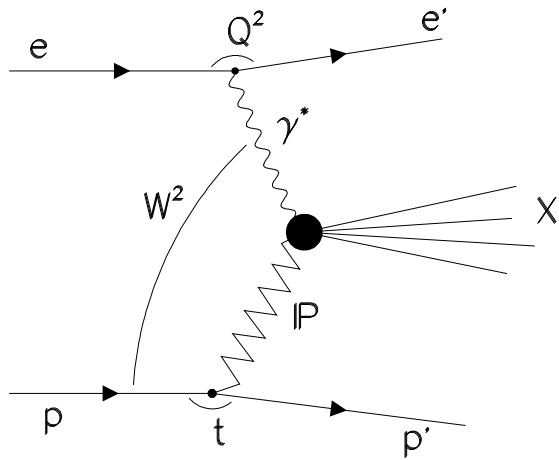}
\caption{Kinematic variables of diffractive $ep$ scattering
at HERA}
\label{kine}
\end{figure}

{\it{HERA Kinematics:}}
The diffractive processes studied at HERA are of the form:
$$ e~(k)~+~ p~(P) \rightarrow e'~(k')~+~p'(P')~+~X,$$
where the photon dissociates into the system $X$ and the outgoing 
proton, $p'$, remains intact, corresponding to single dissociation,
as illustrated in Fig.~\ref{kine}.
The measurements are made as a function of the photon virtuality,
$ Q^2 \equiv -q^2=-(k~-~k')^2, $
the centre-of-mass energy of the virtual-photon proton system,
$W^2=(q~+~P)^2$,
the invariant mass of the dissociated system, $X$, denoted by $M^2$
and the four-momentum transfer at the proton vertex, given by
$t = (P~-~P')^2$.

{\it{Signatures of diffraction:}}
The processes studied build upon the basic elastic scattering process
$$AB \rightarrow AB $$
which is constrained through the measurement of the 
corresponding total cross-section $AB\rightarrow$~anything,
via the optical theorem.
The diffractive processes measured are then
$$AB \rightarrow XB~~~~~~~~  {\rm single~dissociation}$$
$$AB \rightarrow XY~~~~~~~~  {\rm double~dissociation}$$
where the systems $X$ and $Y$ have a limited mass compared to the 
overall energy available, $W$. The process is thought of as being mediated 
by the exchange of an object with vacuum quantum numbers 
(in particular, no colour is exchanged in the process).
A signature for diffraction is via the momentum 
fraction carried by the exchanged colour singlet state. 
When this fraction is less than 1\% of the momentum of, say, 
particle $B$ (in the infinite momentum frame of $B$) we 
can interpret this as being largely due to pomeron exchange.  
The kinematics of producing two low-mass outgoing states 
($X$ and $B$ or $X$ and $Y$)
with a small momentum fraction exchanged between them 
therefore leads to a rapidity gap. 
The high energy available at HERA provides a large rapidity span,
$\Delta(\eta^*)$, of $\sim$ 12 units  
(here, $\Delta(\eta^*) \equiv \eta^*_{\rm max} - \eta^*_{\rm min}$
where $\eta^*_{\rm max} \sim \ln(W/m_p) = 5$ and $\eta^*_{\rm min} \sim
-\ln(W/m_\pi) = -7$ for a typical $W \simeq 150$~GeV).
A colour singlet exchange of reggeons (dominating at lower $W$)
and the pomeron (dominating at higher $W$) can be used to interpret the 
data on single dissociation and double dissociation reactions. 
Fluctuations from processes where colour is 
exchanged may also generate low-mass states. However, these will be
exponentially suppressed: the (Poisson) probability of $not$
producing a given particle in the rapidity gap $\Delta\eta$ when the 
two systems are colour connected is $exp(-\lambda\Delta\eta)$, where
$\lambda$ is the mean particle density for a given $\Delta\eta$ interval.
This exponential fall-off reflects the plateau in the
corresponding (non-diffractive) multiplicity distribution as a function
of $\eta$ which increases relatively slowly (logarithmically) with 
increasing $W$~\cite{nicolo}.

\begin{figure}[htb]
\epsfxsize=10.cm
\centering
\leavevmode
\epsfbox{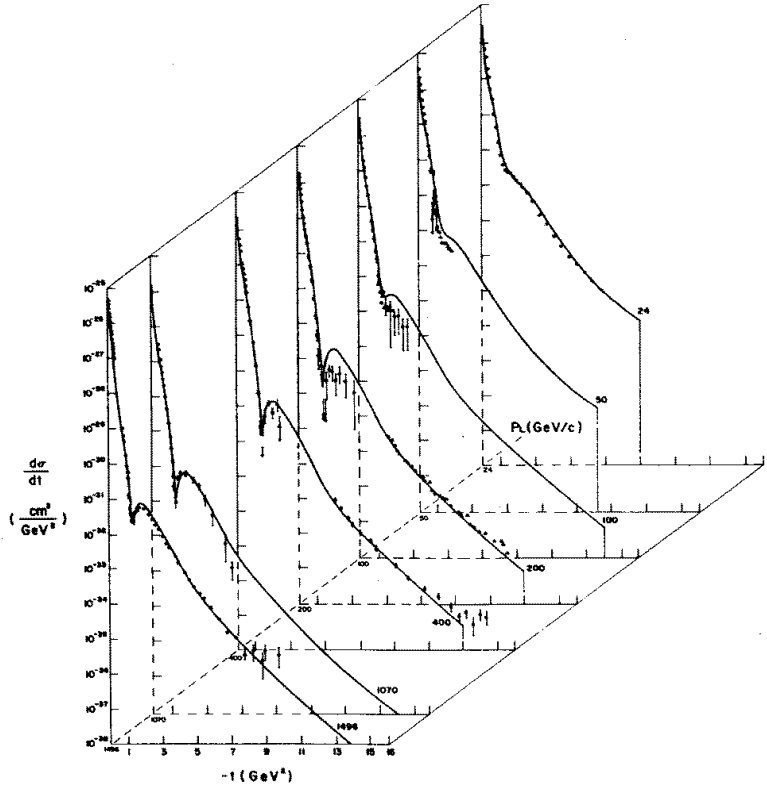}
\caption{{\it Signatures of diffraction:}
$d\sigma/dt$ for ISR $pp$ data as a function of $p_L$, 
the longitudinal momentum of the outgoing proton transformed to 
a fixed-target rest frame, from 24~GeV 
(uppermost plot) to 1496~GeV (lowest plot).}
\label{tsps}
\end{figure}

In Fig.~\ref{tsps}, the $t$ distributions of $pp$ elastic scattering
data are illustrated as a function of the longitudinal momentum ($p_L$)
of the outgoing proton transformed to a fixed-target rest frame.
The patterns
are similar to the diffraction patterns observed when light is scattered 
from an aperture and exhibit an exponential fall-off for values of $t$
below $\simeq$~1~GeV$^2$.  
This characteristic fall-off 
increases with increasing energy, a property known as shrinkage, and 
differs for different incident
and outgoing systems. In order to characterise the $t$-dependence, a fit to
the diffractive peak is performed. 
In the most straightforward approach, 
a single exponential fit to the $t$ distribution, 
$d\sigma/d|t| \propto e^{-b|t|}$ for $|t|~\sleq~0.5~$GeV$^2$ is adopted. 
Physically, the slope of the $t$ dependence in diffractive interactions 
tells us about the effective radius of that interaction, $R_I$: 
if d$\sigma/dt \propto e^{-b|t|}$, then $b \simeq$  $R_I^2/4$. 

{\it{Regge trajectories:}}
The subject of diffraction is far from new: diffractive processes
have been measured and studied for more than thirty years~\cite{goul}. 
Their relationship to the corresponding total cross-sections at high energies
has been successfully interpreted via the optical theorem and Regge theory.
At lower energies the colour singlet exchange of virtual mesons, 
called reggeons, contribute to the fall of the cross-section with increasing
energy. At higher energies,
the introduction of an additional 
trajectory, known as the pomeron trajectory,
with a characteristic $W^2$ and $t$ dependence is 
necessary~\cite{dl}.
The energy behaviour of the total cross-sections can then be described by the
sum of two power-law dependences on the centre-of-mass energy,
$s \equiv W^2$
\begin{eqnarray}
\sigma_{\rm tot} = A\cdot(W^2)^\epsilon + B\cdot(W^2)^{-\eta}
\end{eqnarray}
where $W$ is measured in GeV, 
$\epsilon = \alphapom(0) - 1$ and $\eta$ is defined to be positive
such that $\eta = -(\alpharegge(0)-1)$. Here, 
$\alphapom(0)$ and $\alpharegge(0)$ are the pomeron and reggeon intercepts
(i.e. the values of the parameters at $t$~=~0~GeV$^2$),
respectively.
A wide range of total cross-section data
are used to determine the parameters $\epsilon$ and $\eta$.
The fall-off at low energy due to reggeon exchange 
constrains the value of $\eta \simeq 0.45$.
The slow rise of hadron-hadron total cross-sections with increasing 
energy indicates that the value of $\epsilon \simeq 0.08$ i.e. the total
cross-sections increase as $W^{0.16}$, although the $p\bar{p}$ 
data from CDF at two $\sqrt{s}$ values indicate 
$\epsilon = 0.112 \pm 0.013$~\cite{CDF1}. 
Recent fits using all $pp$ and $\bar{p}p$
data are consistent with a value 
of $\epsilon = 0.08 \pm 0.02$ which will be used here to characterise this 
behaviour~\cite{cudell}.

\begin{figure}[htb]
\epsfxsize=8.cm
\centering
\leavevmode
\epsfbox[28 407 300 650]{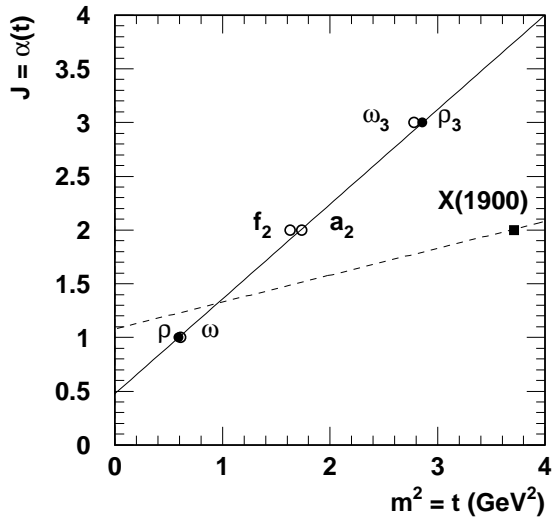}
\caption{{\it Regge trajectories:}
The degenerate regge trajectories are indicated by the solid line.
The pomeron trajectory is indicated by the dashed line.
Also indicated are the $\rho$, $\omega$, $f$ and $a$ resonances as well as
the $I(J^{PC}) = 0(2^{++})$ glueball candidate state $X(1900)$ observed by 
the WA91 collaboration~\protect\cite{WA91}.
}
\label{regge}
\end{figure}         

In a Regge analysis, the diffractive data 
are interpreted via exchanges with spin 
$J = \alpha(t) = \alpha(0) + \alpha't$ and 
$d\sigma/dt \propto (\frac{W^2}{W_0^2})^{2(\alpha(0)-1)} e^{-b|t|}$,
with $b = b_0 + 2\alphappom\ln(W^2/W_0^2)$.
At lower energies, these correspond to reggeon
(i.e. approximately degenerate $\rho$, $\omega$, $f$ and $a$) exchanges.
At the highest energies, where the pomeron contribution dominates,
the optical theorem relates the total cross-sections 
to the elastic, and hence diffractive, scattering amplitude at the same $W^2$.
In Fig.~\ref{regge}, the trajectories, $J = \alpha(t)$, are shown as a function
of $M^2$. 
The diffractive data probe the region of negative $t$. 
Given the dominance of the 
pomeron contribution at large $W$
and an approximately exponential behaviour of the 
$|t|$ distribution with slope $b$, whose mean $|\bar{t}|$ value
is given by $1/b$ at the mean $\bar{W}$ of a given data sample, 
the diffractive cross-section
rise is moderated from $(W^2)^{2\epsilon}$ to 
\begin{eqnarray}
\sigma_{\rm{diff}} \simeq (W^2)^{2(\epsilon-\alphappom\cdot |\bar{t}|)}
\equiv W^{4\bar{\epsilon}}
\end{eqnarray}
where 
$\bar{\epsilon} = \epsilon - \alphappom \cdot |\bar{t}| = \alpha(\bar{t}) - 1 $
and 
$\alphappom = 0.25~$GeV$^{-2}$ reflects the shrinkage of the
diffractive peak as a function of $t$ with increasing $W^2$.
The observed shrinkage of the diffractive peak 
therefore corresponds to a relative reduction
of the diffractive cross-section with increasing energy. 
This value may be compared with the corresponding parameter 
$\alphapregge \simeq 0.9~$~GeV$^{-2}$ for reggeon exchanges.

{\it{Maps of the Pomeron:}}
Whilst these Regge-based models gave a unified description of 
pre-HERA diffractive data, this approach is not fundamentally linked
to the underlying theory of QCD. 
It was anticipated that at HERA and Tevatron energies
if any of the scales $Q^2$, $M^2$ or $t$ become larger than the 
QCD scale $\Lambda^2$, then 
it may be possible to apply perturbative QCD (pQCD) techniques, which 
predict changes to this power law behaviour.
Qualitatively, the $W$ dependence could be ascribed 
to the rise of the gluon density with decreasing $x$ 
determined from the large scaling violations of $F_2(x,Q^2)$,
where $x$ is the Bjorken scaling variable, $x= Q^2/2P \cdot q 
\simeq Q^2/(Q^2+W^2)$.
QCD factorisation into a long-timescale and short-timescale process, 
where this timescale is characterised by $1/Q$ or $1/M$ or $1/\sqrt{t}$,  
leads to the following approaches.
\begin{itemize}
\item[$\bullet$]
For exclusive final states, e.g. vector meson production, with a hard 
scale the approach is very simple
$$\sigma_{\rm{diff}} \sim G_p^2 \otimes \hat\sigma $$
i.e. two-gluon exchange where $G_p^2$ is the square of 
the gluon density of the proton at a representative value of $x$ and
$\hat\sigma$ represents the hard scattering process.
The rise of $F_2$ with decreasing $x$, which constrains the gluon density, 
corresponds to an increase in the effective value of $\epsilon$.
This brings us from the regime of dominance of the slowly-rising
``soft" pomeron to the newly emergent ``hard" behaviour and the question
of how a transition may occur between the two.
The QCD expectation is that the cross-sections should approximately scale as
a function of $t$, corresponding to a weak dependence of $\epsilon$ as a 
function of $t$ and therefore a decrease of $\alphappom$ for the 
perturbative pomeron.
\item[$\bullet$]
For inclusive diffraction with a hard scale
$$\sigma_{\rm{diff}} \sim G_p \otimes \hat\sigma \otimes H$$
i.e. leading-gluon exchange where $G_p$ is the gluon density of 
the proton,
$\hat\sigma$ represents the hard scattering process and $H$ represents
the hadronisation process. Here, the final state with a leading proton 
is seen as a particular hadronisation process~\cite{buch}. 
For processes involving one incoming hadron, the above
approaches can be tested and compared with experimental data.
\item[$\bullet$] 
Finally, one can break the process down further and invoke
Regge factorisation where a flux of pomerons, $f_{\pom/p}$ 
lead to partons from the pomeron, $f_{i/\pom}$, which interact and
hadronise 
$$\sigma_{\rm{diff}} \sim f_{\pom/p} \otimes f_{i/\pom} \otimes 
\hat\sigma \otimes H$$
\end{itemize}
For processes involving two incoming hadrons, this approach has been 
generalised to 
$$\sigma_{\rm{diff}} \sim f_{\pom/p} \otimes f_{i/\pom} 
\otimes \hat\sigma \otimes f_{j/h} \otimes H$$
where $f_{j/h}$ represents the partons from the hadron which
has not diffractively dissociated. This Regge factorisation
approach can therefore be experimentally tested when diffractive data 
from HERA and the Tevatron are compared.

\begin{figure}[htb]
\vspace{1.0cm}
\epsfxsize=10cm
\centering
\leavevmode
\epsfbox[45 497 507 732]{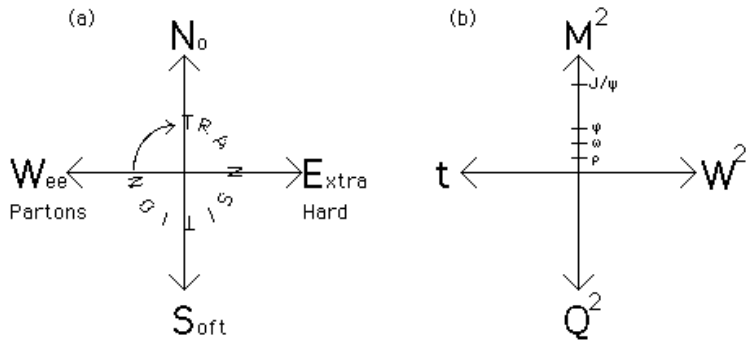}
\caption{{\it{Maps of the pomeron:}} 
(a) theoretical and (b) experimental directions.}
\label{map}
\end{figure}

Precisely where the Regge-based approach breaks down or 
where pQCD may be applicable is open to experimental question.
In addition, once we observe ``hard" diffractive phenomena, we can
ask whether the pQCD techniques applied to inclusive processes 
also apply to these exclusive colour singlet exchange reactions.
The emphasis is therefore on the internal (in)consistency of 
a wide range of measurements of diffractive and total cross-sections.
As an experimentalist navigating around the various theoretical concepts of 
the pomeron, it is sometimes difficult to see which direction to take
and what transitions occur where~(Fig.~\ref{map}(a)).
However, from an experimental perspective, the directions are clear,
even if the map is not yet complete~(Fig.~\ref{map}(b)).

{\it{Outline:}}
The HERA collider allows us to observe a broad range of diffractive 
phenomena at high $W^2$.
What is new is that we have the ability to observe the variation 
of these cross-sections at specific points on the $M^2$
scale, from the $\rho^0$ up to the $\Upsilon$ system as discussed in 
section~2. 
Similarly, the production cross-section can be explored 
as a function of $Q^2$, using a virtual photon probe.
The observation of a significant fraction of events ($\simeq 10\%$)
with a large rapidity gap
between the outgoing proton and the rest of the final state 
in deep inelastic scattering (DIS)
has led to measurements of the internal structure of the pomeron.
In addition, the leading proton spectrometer
data, where the diffracted proton is directly measured, enable 
the $t$ distribution as well as the structure function to be determined 
simultaneously. These results are discussed in section~3.
Studies of the hadronic final state in events with a large
rapidity gap, including transverse energy flows,
event shape distributions and high-$p_T$ jets, 
have been used to provide complementary information on this 
structure.
Also, the observation of rapidity gaps between jets, corresponding to 
large-$t$ diffraction, are presented in section~4.
Many of these hadronic final state investigations were initiated 
at $p\bar{p}$ colliders. In section~5, the latest results from the 
Tevatron on diffractive dijet and $W^\pm$ production, rapidity gaps 
between jets and first observations of hard
double-pomeron exchange are presented. 
A comparison of hard diffractive event rates at HERA and the Tevatron
is given and the interpretation of the observed differences is 
discussed.

\section{Exclusive Production of Vector Mesons}

The experimental signals are the exclusive production of the vector mesons
in the following decay modes
$$\rho^0\rightarrow \pi^+\pi^-~~~~~~
\phi\rightarrow K^+K^-~~~~~
J/\psi\rightarrow\mu^+\mu^-,e^+e^-~~~~~
\Upsilon\rightarrow\mu^+\mu^-.$$
First results on $\omega\rightarrow \pi^+\pi^-\pi^o$ and 
higher vector mesons ($\rho' \rightarrow \pi^+\pi^-\pi^o\pi^o$,
$\rho' \rightarrow \pi^+\pi^-\pi^+\pi^-$ and 
$\psi'\rightarrow\mu^+\mu^-,e^+e^-$) have also been
presented. A recent review of the HERA data can be found in 
Ref~\cite{jim}.

The relevant components of the H1 and ZEUS detectors are the inner tracking
chambers which measure the momentum of the decay products; the calorimeters
which allow identification of the scattered electron and are used in the
triggering of photoproduced vector mesons; and the outer muon chambers used 
to identify muonic decays of the $J/\psi$ and $\Upsilon$.
The clean topology of these events results in typical uncertainties 
on the measured 
quantities ($t$, $M^2$, $W^2$ and $Q^2$), reconstructed in the tracking 
chambers, of order 5\%.
Containment within the tracking chambers corresponds to a $W$ interval
in the range $40 \sleq W \sleq 180$~GeV. 

Photoproduction processes have been extensively studied in 
fixed-target experiments,
providing a large range in $W$ over which to study the cross-sections.
The key features are the weak dependence of the cross-section on $W$, 
an exponential dependence on $t$ with a $b$ 
slope which shrinks with increasing $W$
and the retention of the helicity of the photon by the vector meson. 

\begin{figure}[htb]
\epsfxsize=10.cm
\centering
\leavevmode
\epsfbox{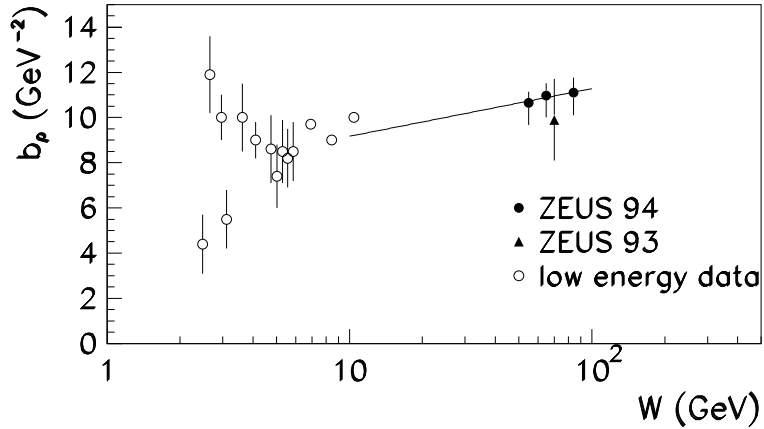}
\caption{ZEUS exclusive $\rho^0$ $t$ distributions characterised by $b_\rho$ 
as function of $W$ together with a fit to the data discussed in the text
and compared to a compilation of lower energy data~\protect\cite{aston}.}
\label{trho}
\end{figure}

In Fig.~\ref{trho}, the ZEUS results for exclusive $\rho^0$ 
production as a function of $t$ are shown.
An exponential fit to the ZEUS data in three $W$ intervals yields $b$-slopes
which are fitted to the form
$$b = b_0 + 2\alphappom\ln(W^2/M_V^2)$$
expected from Regge theory.
The fitted value of $\alphappom = 0.23 \pm 0.15 ^{+ 0.10}_{-0.07}$~GeV$^{-2}$
is consistent with a shrinkage of the $t$ slope with 
increasing $W^2$ expected from soft diffractive processes.

\begin{figure}[htb]
  \centering
\mbox{
\subfigure[$b$-slopes as a function of $M_V^2$.]
{\psfig{figure=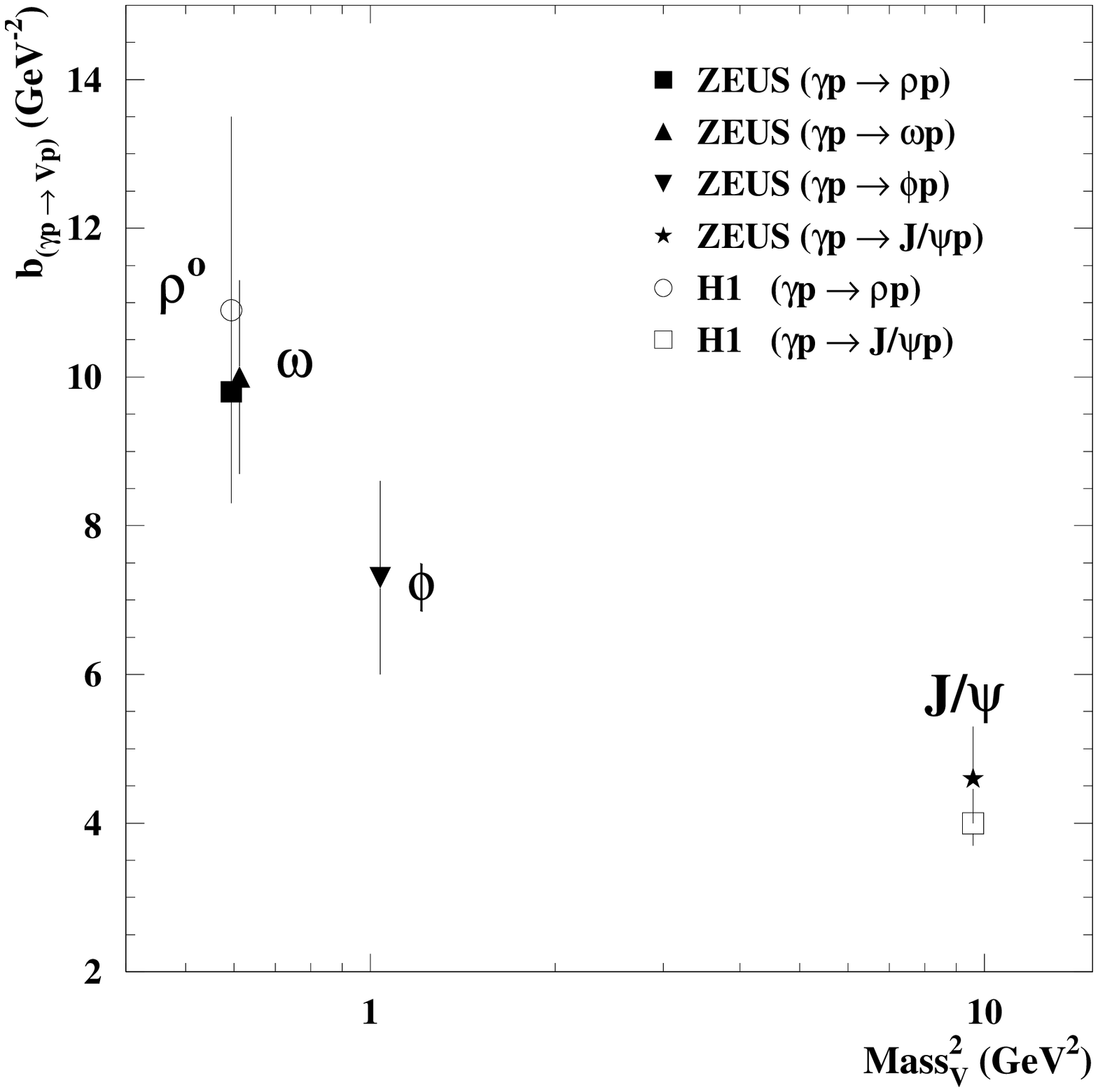,width=.41\textwidth}}\quad
\subfigure[$b$-slopes as a function of $<Q^2>$ for exclusive $\rho^0$
production.]
{\psfig{figure=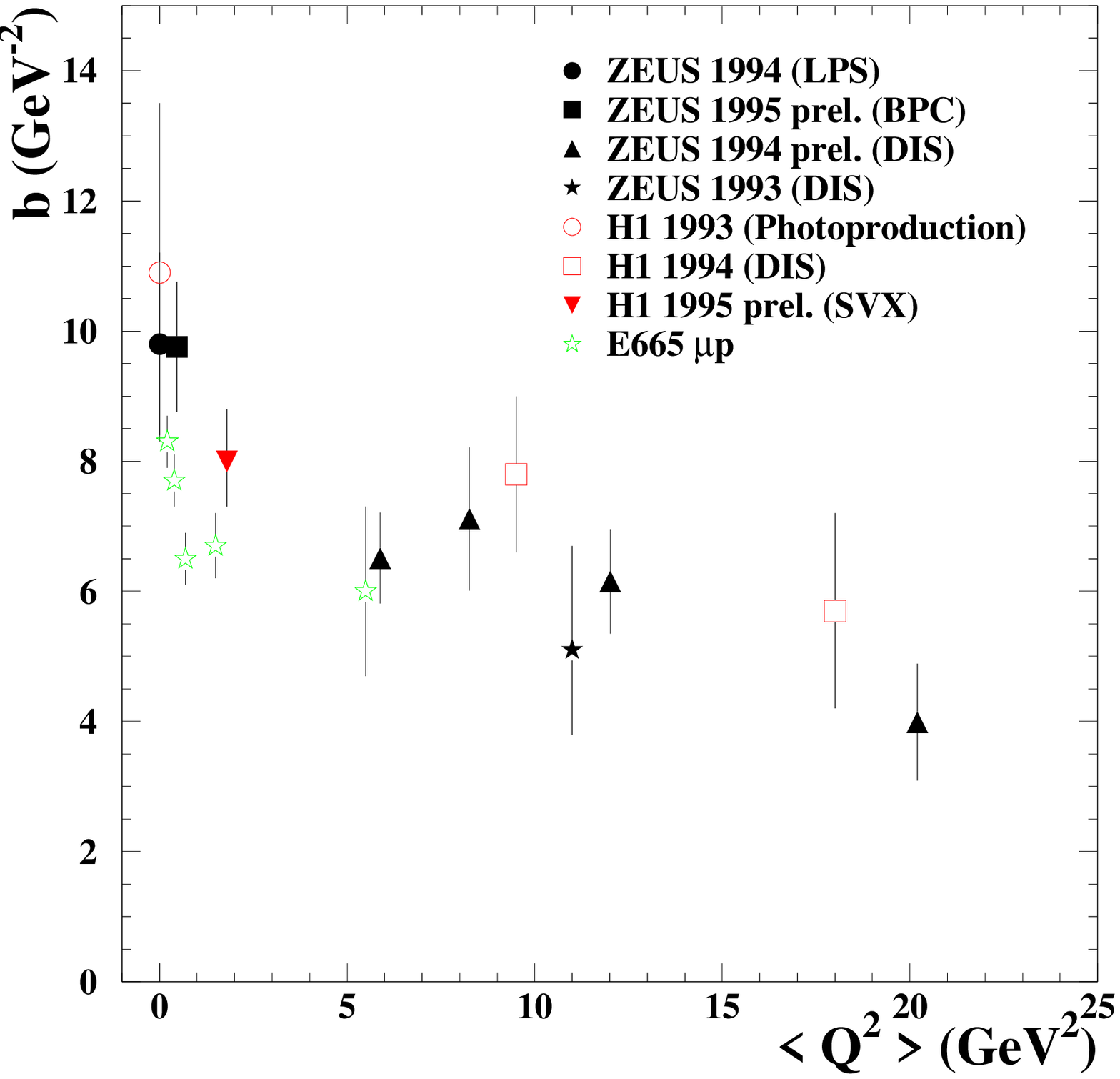,width=.45\textwidth}}
}
  \caption[]{Exclusive vector meson production $b$-slopes as a function of
(a) mass of the vector meson $M_V^2$ and (b) mean virtuality of the photon 
$<Q^2>$.}
\label{bmq2}
\end{figure}

The interaction radius, $R_I$, can be approximately related to the
radii of the interacting proton and vector meson, 
$R_I \simeq \sqrt{R_P^2 + R_V^2}$.
The variation of these $b$ values is shown in Fig.~\ref{bmq2}(a) 
as a function of
vector meson mass $M_V^2$.
In Fig.~\ref{bmq2}(b), these slopes are presented as a function of increasing 
virtuality of the photon for $\rho^0$ production data.
In each case, the range of measured $b$-slopes varies from 
around 10~GeV$^{-2}$ ($R_I \simeq 1.3$~fm) at low $M_V^2$ or $Q^2$ to  
4~GeV$^{-2}$ ($R_I \simeq 0.8$~fm) at the highest $M_V^2$ or $Q^2$ so far 
measured.
Given $R_P \simeq 0.7$~fm, 
this variation in $b$-slopes corresponds to a significant decrease 
in the effective radius of the interacting vector meson from 
$R_V \simeq 1.1$~fm to $R_V \simeq 0.4$~fm as $M_V^2$ (at fixed $Q^2 \simeq
0$~GeV$^2$) or $Q^2$ (at fixed $M_V^2 = M_\rho^2$) increase.

\begin{figure}[htb]
\epsfxsize=12.cm
\centering
\leavevmode
\epsfbox{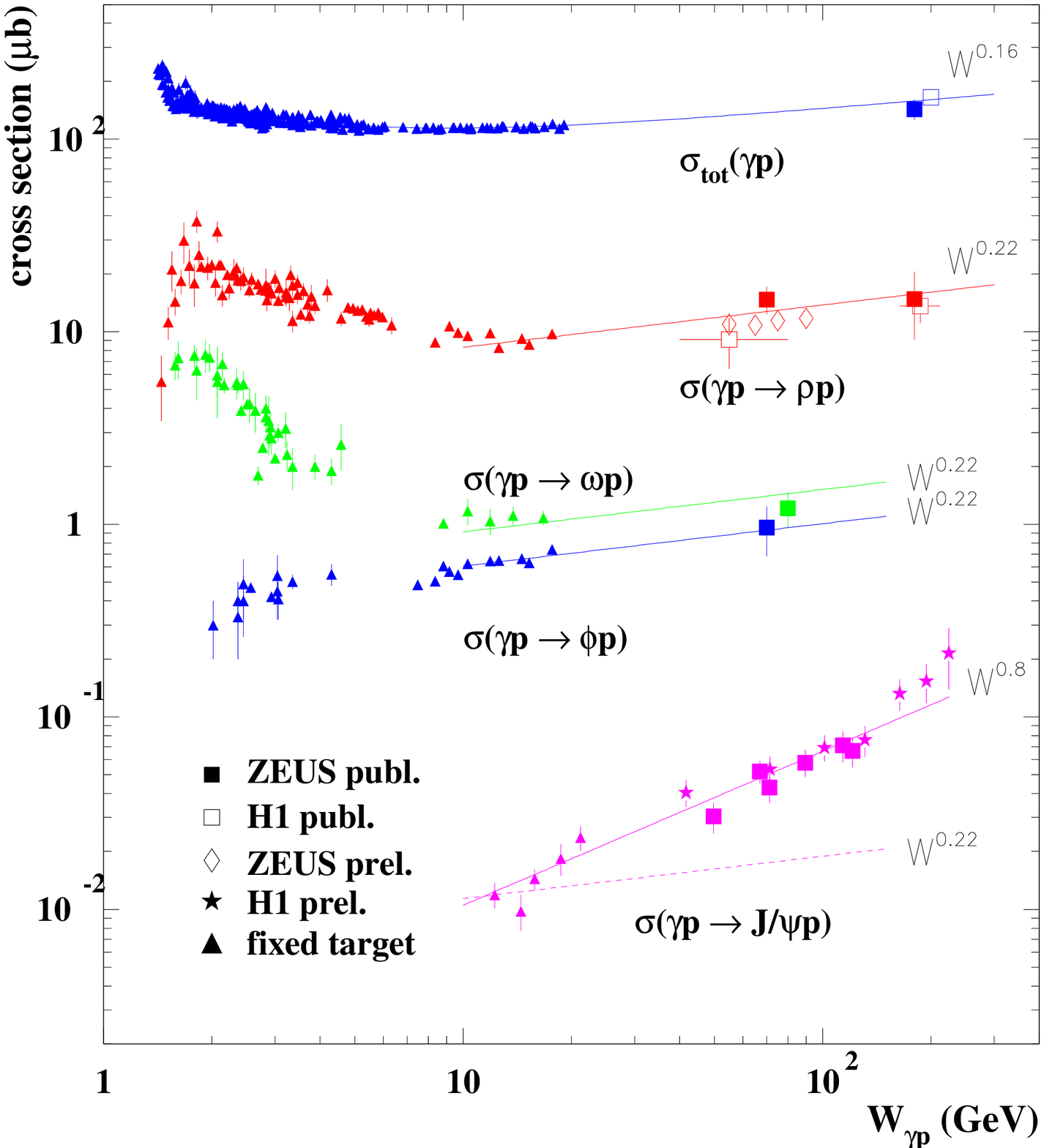}
\caption{$W$ dependence of the exclusive vector meson and total 
photoproduction cross-sections compared to various power law dependences
discussed in the text.}
\label{wvm}
\end{figure}         

Integrating over the measured $t$ dependence, 
the $W$ dependence of the results on 
exclusive vector meson photoproduction cross-sections 
are shown in Fig.~\ref{wvm}. There is generally good agreement 
between the experiments on the measured cross-sections.
The $\gamma p$ total cross-section is also shown in Fig.~\ref{wvm}, 
rising with increasing 
energy as in hadron-hadron collisions and consistent with a value of 
$\epsilon \simeq 0.08$ i.e. the total cross-section increases as $W^{0.16}$. 

Given the dominance of the pomeron trajectory at high $W$ and
a $|t|$ distribution whose mean value $|\bar{t}|$ 
is given by $1/b$, the diffractive cross-section
rise is moderated from
$W^{4\epsilon}= W^{0.32}$
to 
$$W^{4(\epsilon - \alphappom\cdot|\bar{t}|)} 
\equiv W^{4\bar{\epsilon}} = W^{0.22}.$$
Here $\bar{\epsilon} = 0.055$ characterises the effective energy dependence 
after integration over $t$ for
$b = 10$~GeV$^{-2}$ (which is appropriate for $\rho^0$ exchange as observed 
in Fig.~\ref{trho}).
The observed shrinkage of the diffractive peak 
therefore corresponds to a relative reduction
of the diffractive cross-section with increasing energy. 
Such a dependence describes the general increase of the 
$\rho^0$, $\omega$ and $\phi$ vector meson cross-sections with increasing $W$.
However, the rise of the $J/\psi$ cross-section is clearly not described 
by such a $W$ dependence, the increase being described by an 
effective $W^{0.8}$ dependence. Whilst these effective powers are for
illustrative purposes only, it is clear that in exclusive $J/\psi$ production
a new phenomenon is occurring. For example, 
fits to the ZEUS data yield results for
$\bar{\epsilon}$ which are inconsistent with soft pomeron exchange
$$\bar{\epsilon} = 0.230 \pm 0.035 \pm 0.025.$$

\begin{figure}[htb]
  \centering
\mbox{
\subfigure[$W$ dependence of exclusive $J/\psi$ 
photoproduction cross-sections 
compared to the Ryskin model discussed in the text.]
{\psfig{figure=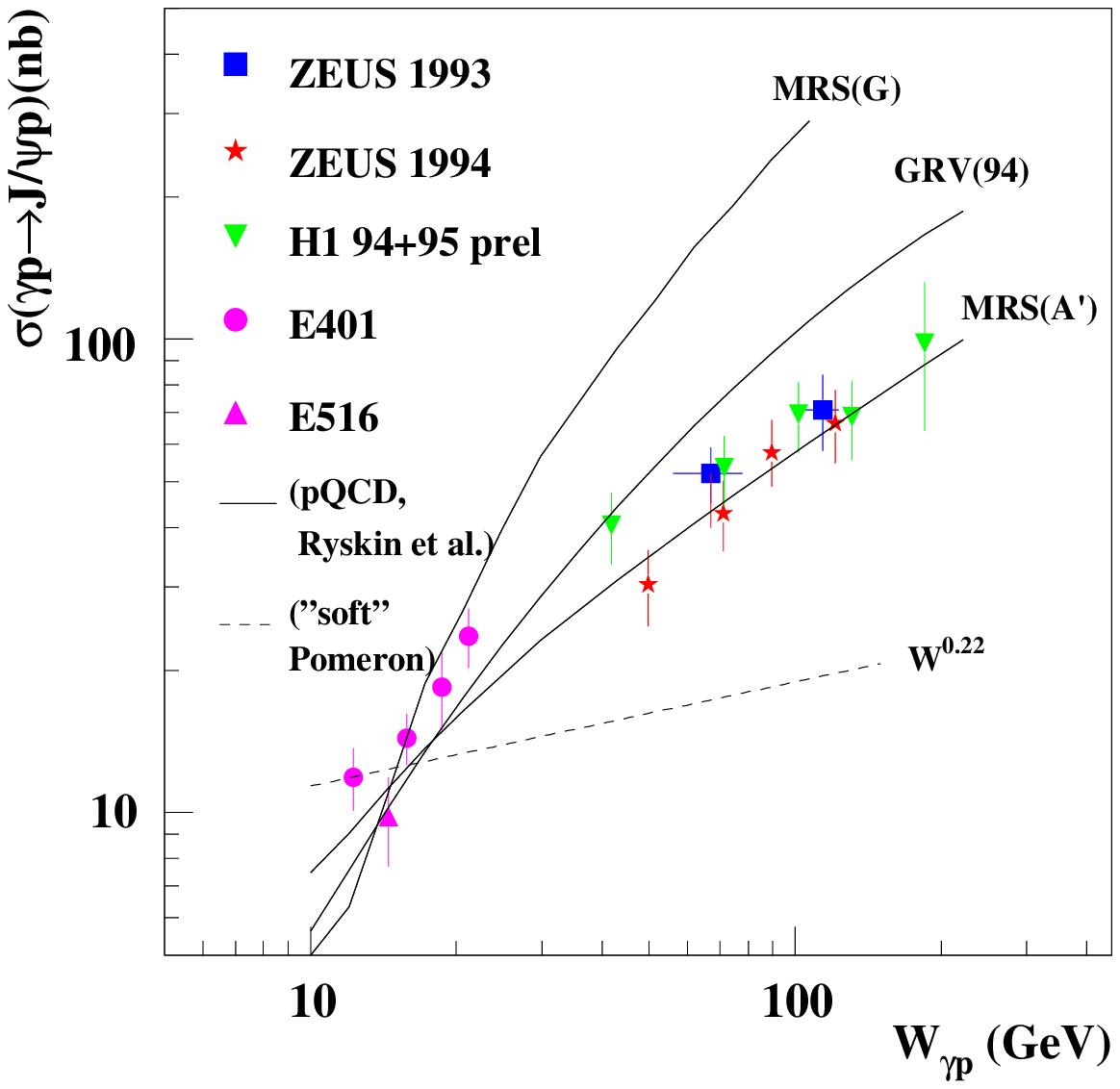,width=.45\textwidth}}\quad
\subfigure[$Q^2+M_\psi^2$ dependence of exclusive $J/\psi$ electroproduction
cross-sections compared to the dependences discussed in the text.]
{\psfig{figure=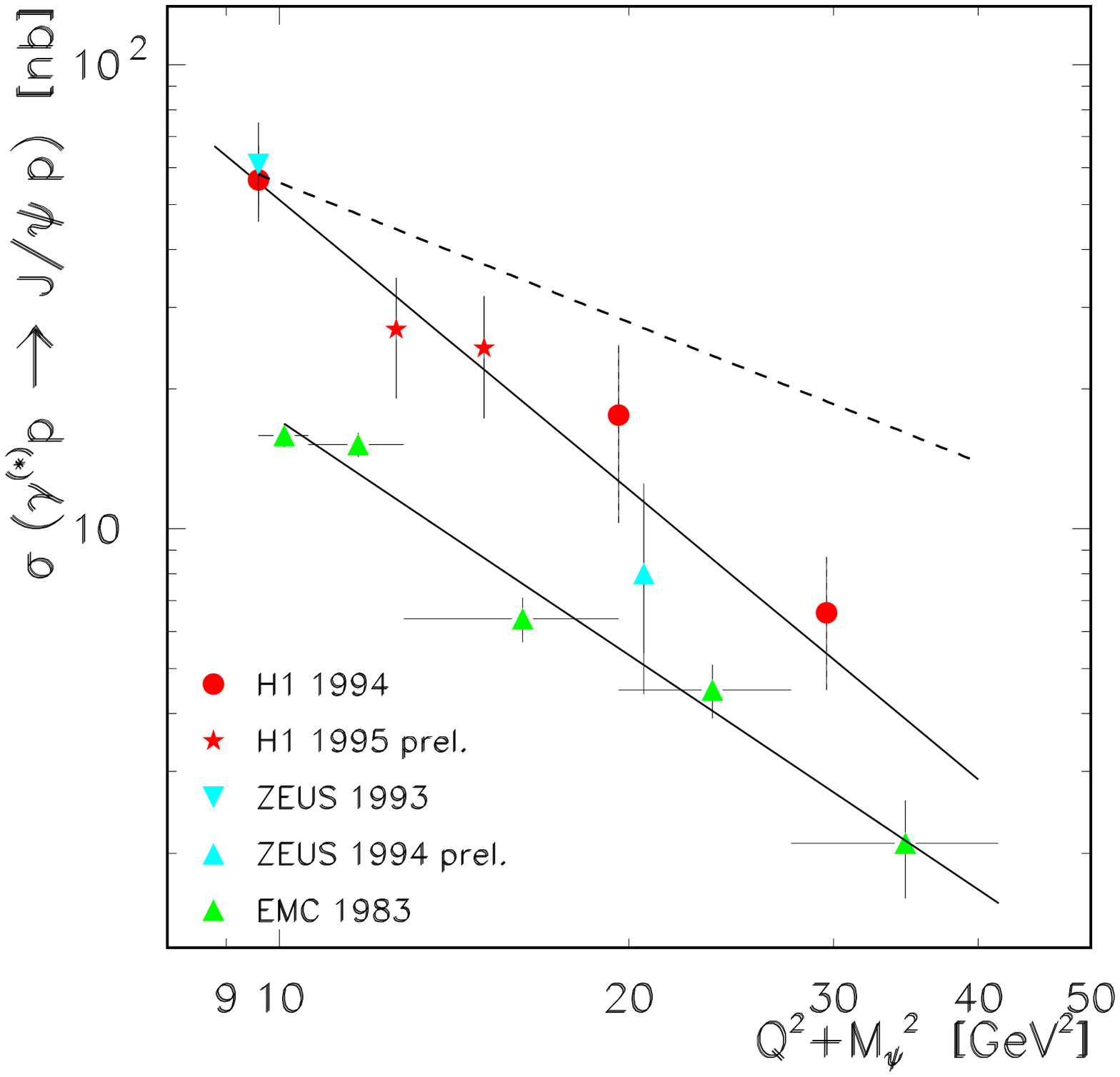,width=.45\textwidth}}
}
  \caption[]{Exclusive $J/\psi$ production cross-sections as a function of
(a) $W$ and (b) $Q^2+M_\psi^2$.}
\label{jpsiwq2}
\end{figure}

The $J/\psi$ (charm) mass scale, $M_\psi^2$, is larger than the QCD scale, 
$\Lambda^2$, and it is therefore possible to apply pQCD 
techniques.
The theoretical analysis predicts that 
the rise of the cross-section is proportional to the square of the gluon
density at small-$x$ (a pair of gluons with no net colour
is viewed as the perturbative pomeron)
$$\sigma (\gamma p \rightarrow J/\psi~p) \propto [(xG(x,\bar{Q^2})]^2 
\simeq W^{4\bar{\epsilon}} \simeq W^{0.8}$$
where $\bar{Q^2}= (Q^2 + M_{J/\psi}^2)/4 (\simeq 2.4$~GeV$^2$
for the photoproduction data) 
is chosen as the scale in the Ryskin model~\cite{ryskin}.
This approach enables discrimination among recent 
parametrisations of the proton structure function as shown in 
Fig.~\ref{jpsiwq2}(a). Here, the measured cross-sections are
compared to the Ryskin model for various choices of parton densities 
which describe recent $F_2$ data.
The approach is therefore very promising as an independent 
method to determine 
the gluon distribution at low-$x$ from the HERA data.
Currently, however, there are model uncertainties and the calculations 
are only possible in leading order. The normalisation is therefore 
uncertain by up to a factor of two.

We also know from measurements of the DIS $\gamma^* p$ total cross-section 
that application of formula~(1) results in a value of $\epsilon$ 
which increases with increasing $Q^2$, with $\epsilon \simeq$ 0.2 to 0.25 
at $Q^2\simeq10$~GeV$^2$~\cite{levy}.
The fact that the corresponding relative rise of $F_2$ with decreasing 
$x$ can be described by pQCD evolution~\cite{GRV}
points towards a predicted function $\epsilon = \epsilon(Q^2)$
for $Q^2 \sgeq Q_o^2$. The current data indicate that 
this transition occurs for $Q^2 \simeq 1$~GeV$^2$~\cite{zeusf2}.

$J/\psi$ electroproduction results are also available;
the $Q^2+M_\psi^2$ dependence of the data 
is shown in Fig.~\ref{jpsiwq2}(b). The cross-sections are fitted to 
$\sigma \propto 1/(Q^2+M_\psi^2)^n$ where the fitted line 
corresponds to $n \simeq 2.1$ for the combined H1 and ZEUS data.
This compares to the prediction of the Vector Dominance Model (VDM),
applicable to soft photoproduction processes, where $n$~=~1 (shown as the
dashed line) and the 
Ryskin model where $n~\simeq~3$. 
Also shown are the lower-$W$ EMC data where $n \simeq 1.5$.
The $J/\psi$ electroproduction cross-section is of the same order as 
the $\rho^0$ data. This is in marked contrast to the significantly lower 
photoproduction cross-section for the $J/\psi$, even at HERA energies, 
also shown in Fig.~\ref{wvm}.
Further results in this area will allow tests of the underlying dynamics 
for both transverse and longitudinally polarised photons coupling to light
and heavy quarks in the pQCD calculations.

\begin{figure}[htb]
\epsfxsize=14.cm
\centering
\leavevmode
\epsfbox{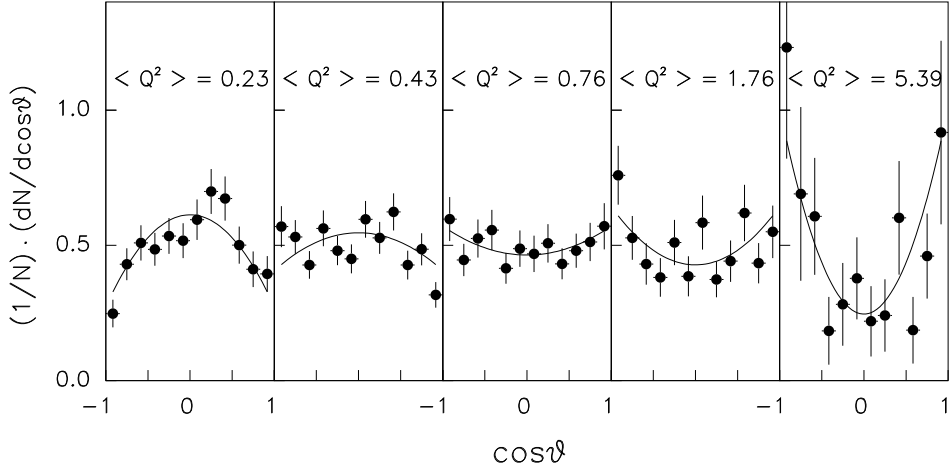}
\caption{$Q^2$ dependence of the decay angular distributions for E665 
$\rho^0$ electroproduction data.}
\label{q2ang}
\end{figure}         

\begin{figure}[hbt]
\epsfxsize=8.cm
\centering
\leavevmode
\epsfbox{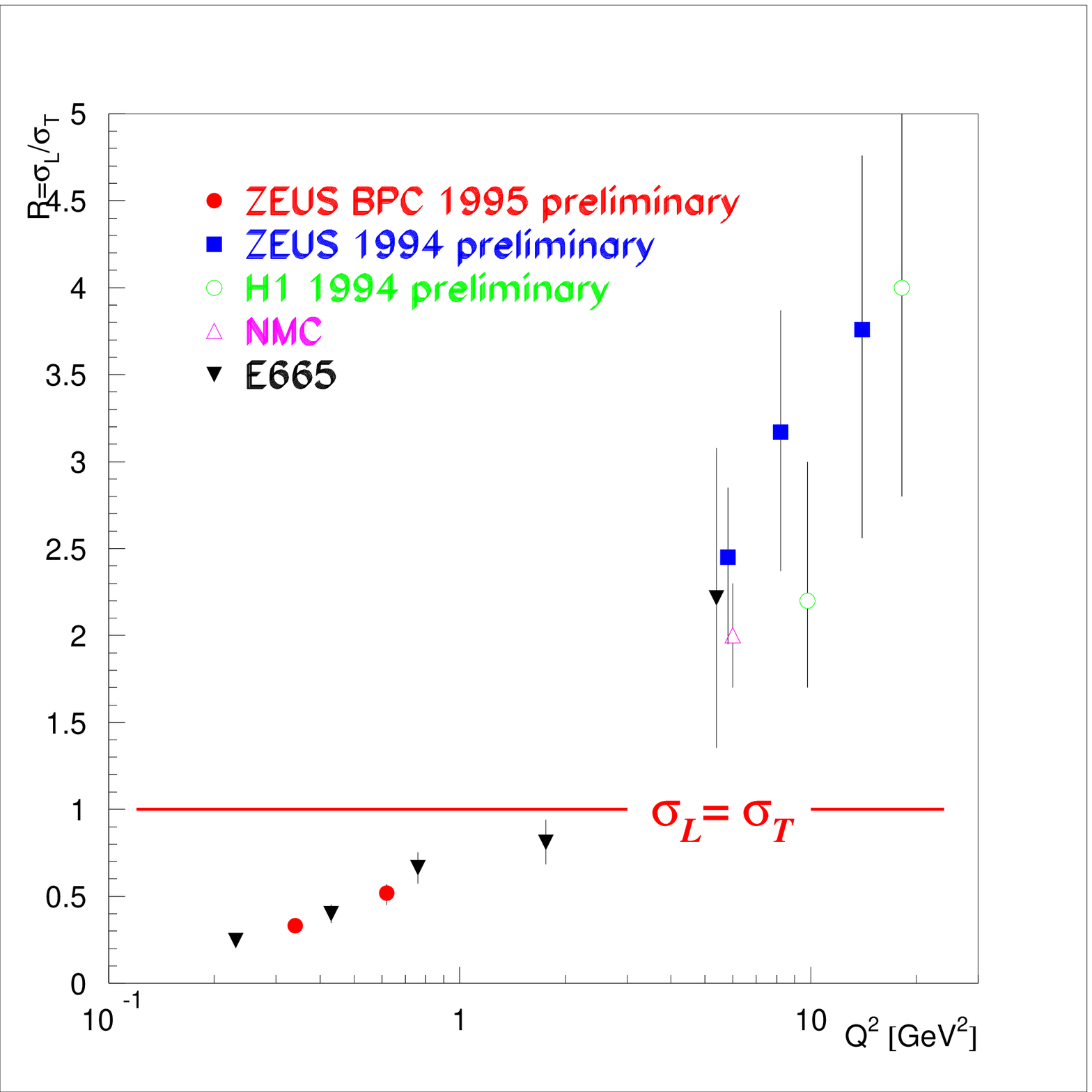}
\caption{Ratio $R= \sigma_L/\sigma_T$ of $\rho^0$
electroproduction data.}
\label{Rplot}
\end{figure}         

One contribution to the DIS $\gamma^* p$ total cross-section is
the electroproduction of low mass vector mesons, here typified by the 
$\rho^0$ data. 
The decay angle distributions 
of the pions in the $\rho^0$ rest frame with respect to the 
virtual photon proton axis from the E665 fixed-target experiment ($7<W<28$~GeV)
are shown in Fig.~\ref{q2ang}~\cite{MPI}.
The measurements of this helicity angle of the vector meson decay 
determines 
$R= \sigma_L/\sigma_T$ for the (virtual) photon, assuming 
$s$-channel helicity conservation, i.e. 
that the vector meson preserves the helicity of the photon.

The decay angular distribution can be written as
$    \frac {1}{N}\frac{dN}{d\rm{cos} \it \theta_h} =
   \frac{3}{4}[1-r_{00}^{04}+(3 r_{00}^{04}-1)\rm{cos}^2\it \theta_{h}],
$
where the density matrix element $r_{00}^{04}$ represents
the probability that the $\rho^0$ was produced 
longitudinally polarised by either transversely
or longitudinally polarised virtual photons.
The value of $R$ is then obtained from
$     R = \frac{\sigma_L}{\sigma_T} = 
\frac{2(1-y)}{Y_+}\cdot \frac{r_{00}^{04}}{1-r_{00}^{04}}$,
where $y = P \cdot q/P \cdot k$ 
is the fractional energy loss of the electron in the
proton rest frame and $Y_+ = 1 + (1-y)^2$. The 
kinematic factor $2(1-y)/Y_+$ is typically close to unity.
This variation of $R$ with $Q^2$ is summarised in Fig.~\ref{Rplot}.
The photoproduction measurements for the $\rho^0$ (not shown)
are consistent with the interaction of dominantly transversely polarised 
photons and hence $R \simeq 0$. 
The electroproduction data are consistent with a universal
dependence on $Q^2$ independent of $W$ and show a transition from 
predominantly transverse to predominantly longitudinal photons
with increasing $Q^2$. 
This increase of $\sigma_L$ is due to an increased flux of longitudinal
photons, $\sigma_L \propto (Q^2/M_X^2) \sigma_T$. 
At higher $Q^2$ values, the cross-section due to longitudinal exchange
is determined in leading-log pQCD~\cite{brodsky} where the underlying 
interaction of the virtual photon with the constituent quarks of the 
$\rho^0$ is calculated.
As noted previously (see Fig.~\ref{bmq2}(b)), the measured 
$b$-slope decreases by about a factor of two
from the photoproduction case to values comparable to that
in the photoproduced $J/\psi$ case.
The basic interaction is probing smaller distances, 
which allowed a first comparison
of the observed cross-section with the predictions of 
pQCD~\cite{Zrho*}.

\begin{figure}[htb]
  \centering
\mbox{
\subfigure[World data on exclusive $\rho^0$ production.]
{\psfig{figure=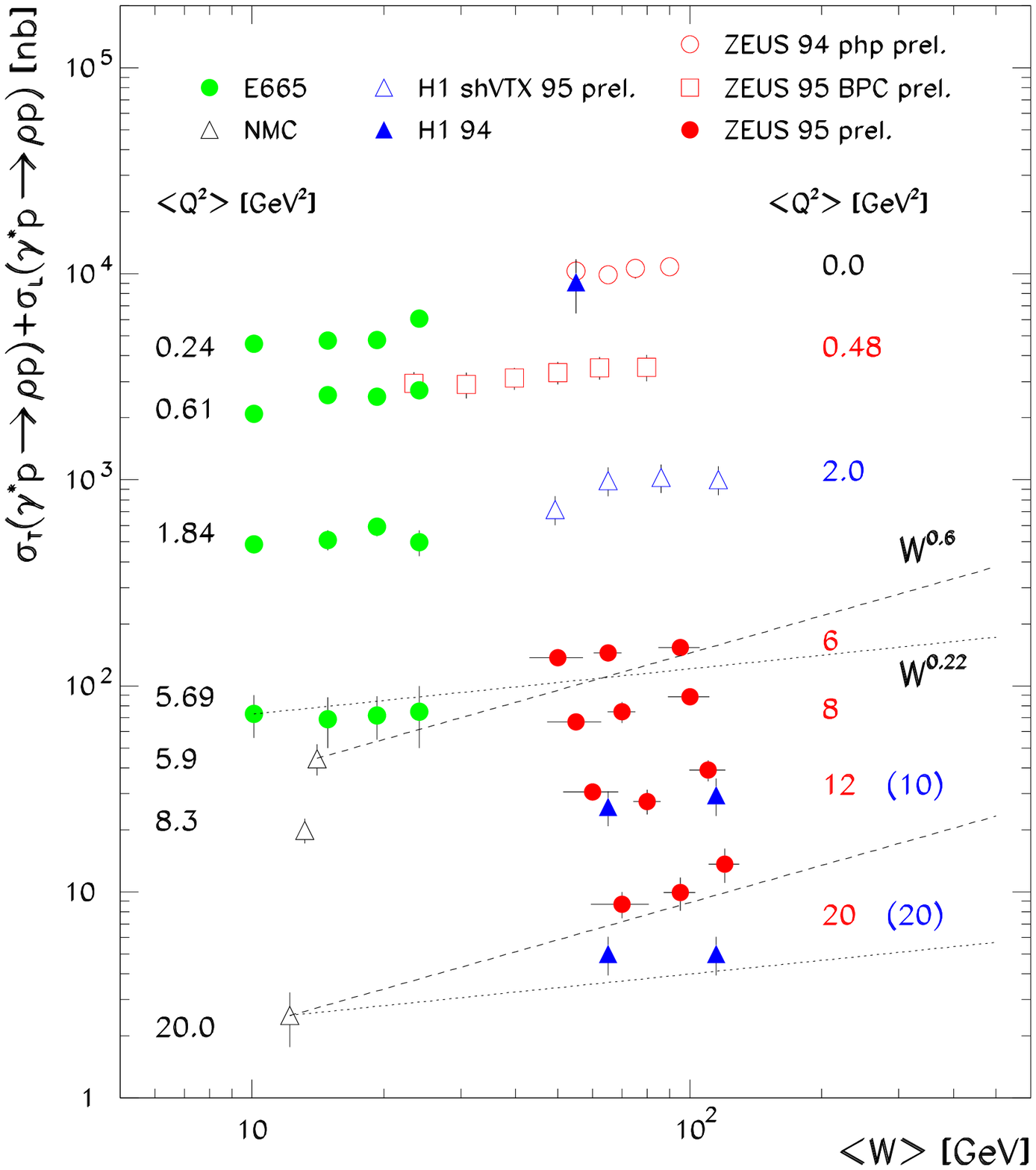,width=.45\textwidth}}\quad
\subfigure[ZEUS exclusive $\rho^0$ electroproduction data.]
{\psfig{figure=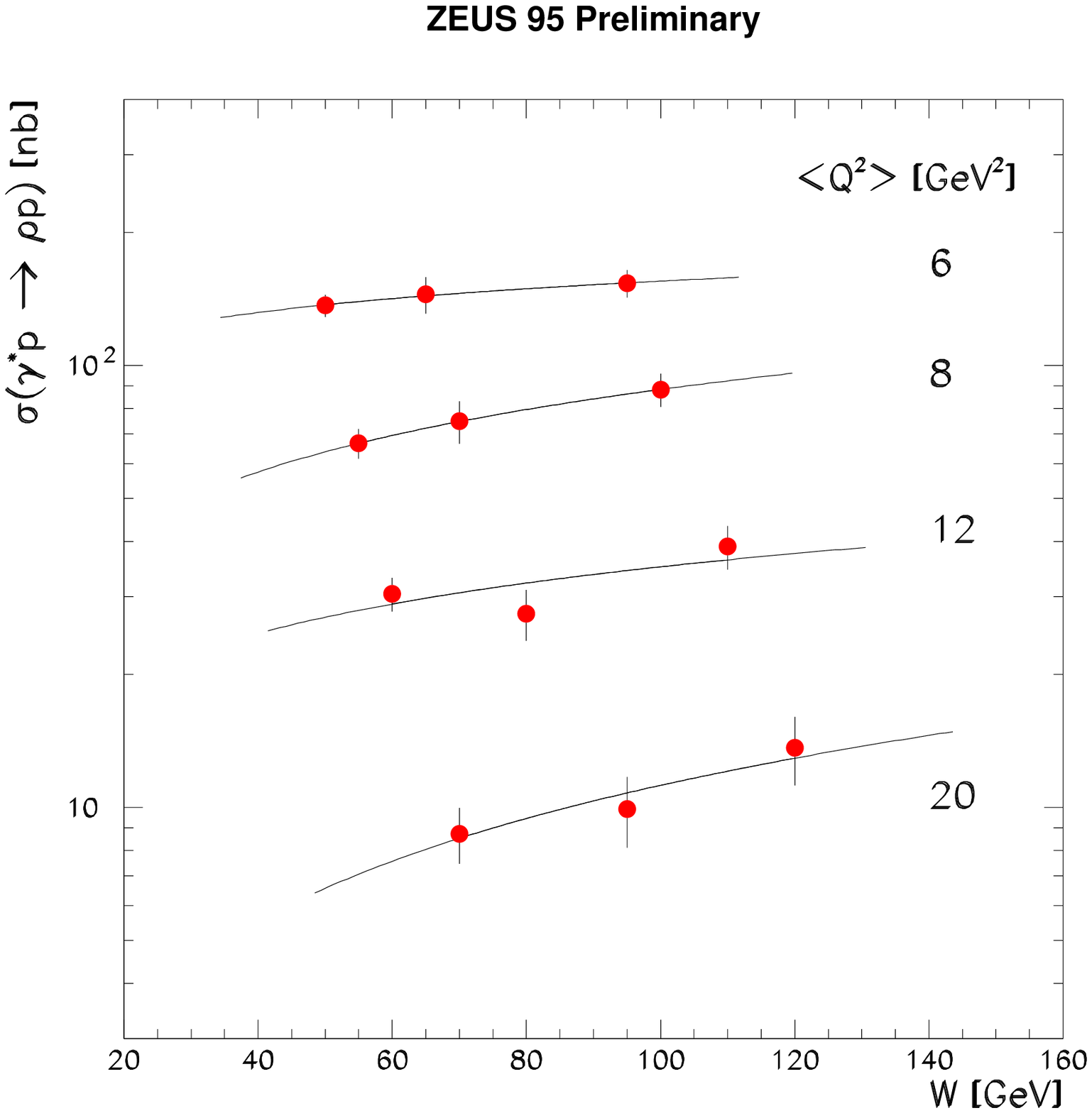,width=.45\textwidth}}
}
  \caption[]{Exclusive $\rho^0$ virtual-photon proton cross-sections as 
a function of $W$ for (a) all data and (b) ZEUS preliminary data, compared 
to the dependences discussed in the text.}
\label{rhoq2}
\end{figure}

The $W$ dependence of the (virtual-)photon proton $\rho^0$ 
cross-sections for finite values of $Q^2$ are shown in Fig.~\ref{rhoq2}(a),
compared to the corresponding photoproduction cross-sections
(the $\phi$ data, not shown, exhibit similar trends).
There is a significant discrepancy between the ZEUS  and H1 measured
cross-sections at $Q^2 = 20$~GeV$^2$ as well as a smaller discrepancy between 
the E665 and NMC measurements at $Q^2 \simeq 6$~GeV$^2$. This is illustrated by
comparison with the $W^{0.8}$ (dashed) and $W^{0.22}$ dotted lines for 
$Q^2 \simeq 6$~GeV$^2$ and $Q^2 \simeq 20$~GeV$^2$. At each $Q^2$ value,
a simple dependence cannot account for all the data.

One of the key problems in obtaining accurate
measurements of these exclusive cross-sections and the $t$ slopes 
is the uncertainty of the double dissociation component, where the 
proton has also dissociated into a low mass nucleon system~\cite{dd}. 
At HERA, the forward calorimeters will see the dissociation
products of the proton if the invariant mass of the nucleon system, $M_N$,
is above approximately 4~GeV. 
A significant fraction of double dissociation events produce a limited mass 
system which is typically not detected.
One expects that the dissociated mass spectrum will fall 
as $ d\sigma/dM_N^2 \propto (1/M_N^2)^{1+\bar{\epsilon}} $
and integrating over the $t$-dependence CDF obtains
$\epsilon$ = 0.100 $\pm$ 0.015 at $\sqrt{s} = 1800$~GeV~\cite{CDF2}.
Precisely how the proton dissociates and to what extent the proton
can be regarded as dissociating independently of the photon system is not 
$a~priori$ known. 
Currently, this uncertainty is reflected in the cross-sections
by allowing the value of $\bar{\epsilon}$ to vary from around 0.0 to 0.5, 
a choice which covers possible variations of $\bar{\epsilon}$ as a function
of $M_N$ and $W$.
Combining all uncertainties, the overall systematic errors on the 
various cross-sections are typically $\simeq~20\%$ for both 
the photoproduction and electroproduction measurements.
The estimation of the double dissociation contribution has, however, 
historically 
been one of the most significant experimental problems with these measurements.
Whether this is the source of the H1 and ZEUS discrepancy is not yet known.

The combined $W$ dependence of the $\rho^0$ 
electroproduction data are, therefore, currently inconclusive.
However, taking the ZEUS data alone, shown in Fig.~\ref{rhoq2}(b)
there are indications of a transition from
the soft to the hard intercept with $\bar{\epsilon}$ varying from
$ \bar{\epsilon} = 0.04 \pm 0.01 \pm 0.03~(Q^2 = 0.5~{\rm GeV}^2)$
to
$ \bar{\epsilon} = 0.19 \pm 0.05 \pm 0.05~(Q^2 = 20~{\rm GeV}^2)$
as indicated by the fitted lines in Fig.~\ref{rhoq2}.
These data are therefore consistent
with a $W^{0.22}$ ($\bar{\epsilon}$ = 0.05) 
dependence at the lowest $Q^2$ values and the $W^{0.8}$ 
($\bar{\epsilon}$ = 0.2) dependence at the highest $Q^2$ values.

\begin{figure}[hbt]
\epsfxsize=8.cm
\centering
\leavevmode
\epsfbox[27 178 523 676]{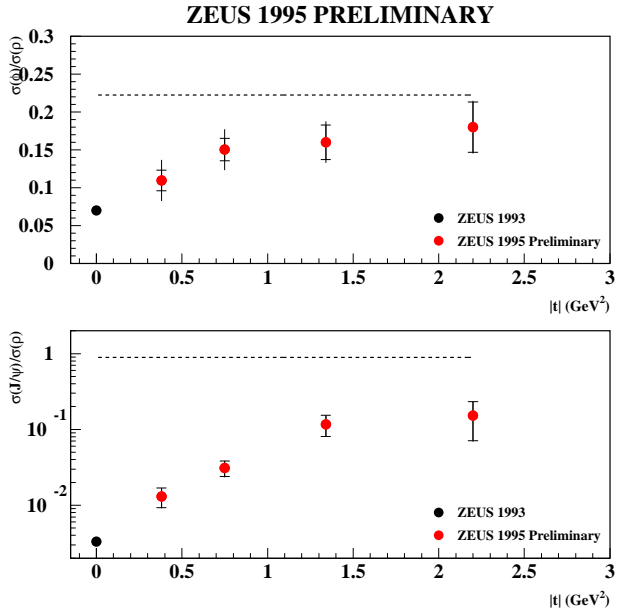}
\caption{$t$ dependence of the ratio of exclusive production cross-sections 
$\sigma(\phi)/\sigma(\rho^0)$ (upper plot) and
$\sigma(J/\psi)/\sigma(\rho^0)$ (lower plot) 
for the ZEUS photoproduction data.}
\label{tratio}
\end{figure}

An important point to emphasise here is that the relative production of 
$\phi$ to $\rho^0$ mesons approaches the quark model prediction of 2/9 
at large $W$ as a function of $Q^2$.
Similar observations have been made on the $t$ dependence
of this ratio for photoproduction data, as shown in the upper
plot of Fig.~\ref{tratio}.
Here, the ratio of the $\phi/\rho^0$ cross-sections approaches the SU(4)
flavour prediction of $\rho:\omega:\phi:J/\psi = 
[\frac{1}{\sqrt{2}}(u\bar{u} - d\bar{d})]^2:
[\frac{1}{\sqrt{2}}(u\bar{u} + d\bar{d})]^2:
[s\bar{s}]^2:
[c\bar{c}]^2 = 9:1:2:8$.
The restoration of this symmetry indicates that the photon is interacting
via quarks, rather than as a vector meson with its own internal structure.
This therefore indicates the relevance of a gluonic interpretation of the 
pomeron and the applicability of pQCD to these cross-sections.
Similarly, 
the relative production of $J/\psi$ to $\rho^0$ mesons
is shown with the asymptotic prediction of 8/9 from the quark model
in the lower plot of Fig.~\ref{tratio}.
In this case, it is evident that threshold effects for the heavy charm
quark are still significant in the measured $t$ range,
however the ratio climbs by almost two orders of magnitude 
from $t\simeq 0$ to $t\simeq 2$~GeV$^2$. 

\begin{figure}[hbt]
\epsfxsize=7.cm
\centering
\leavevmode
\epsfbox[18 142 575 700]{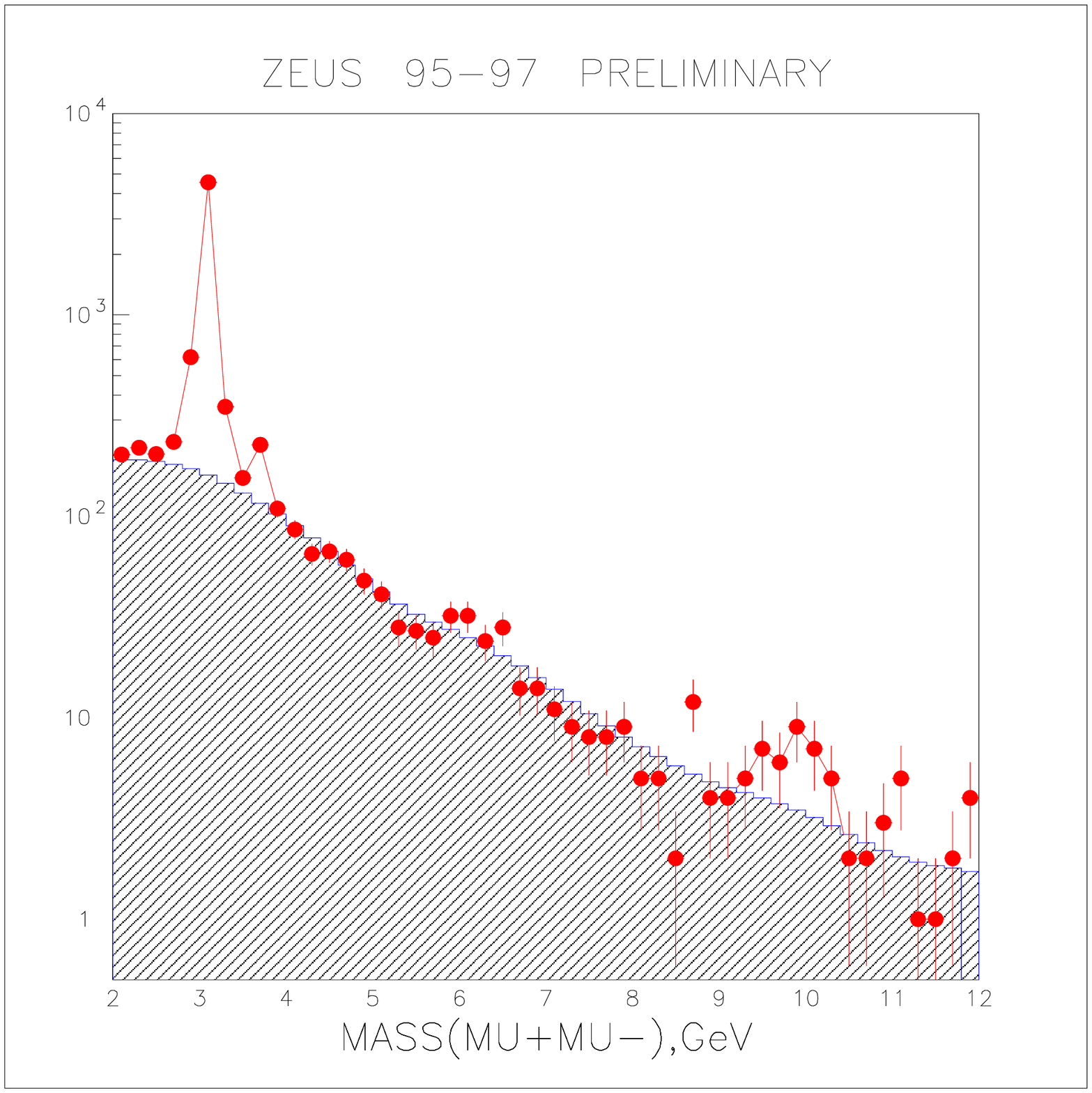}
\caption{Observation of exclusive $\Upsilon$ production: the 
invariant mass spectrum for $\mu^+\mu^-$ indicates a broad enhancement 
around 10~GeV above the fitted background.}
\label{upsm}
\end{figure}

Exclusive $\Upsilon$ photoproduction has also been observed~\cite{eg}. 
The $\simeq$ 20 events observed in the $\mu^+\mu^-$ channel 
corresponds to a cross-section for predominantly (1S) production, 
as well as higher $\Upsilon$ states, 
of $0.9\pm0.3\pm0.3$~nb for 
$\sigma (\gamma p \rightarrow \Upsilon N)$ where $M_N <$~4~GeV
and $80<W<280$~GeV, $Q^2 < 4$~GeV$^2$. This 
is about 1\% of the $J/\psi$ cross-section, as shown in 
Fig.~\ref{upsm},
emphasing the importance of the mass of the heavy quark in the
production of exclusive vector mesons.

In conclusion, there is an accumulating body of exclusive vector meson
production data, measured with a systematic precision of $\simeq 20\%$, which
exhibit two classes of $W^2$ behaviour: a slow rise consistent with that
of previously measured diffractive data for low $M_V^2$ photoproduction
data but a significant rise of these cross-sections 
above a finite value of $M_V^2$, $t$ or $Q^2$.
In general, the cross-sections at large $W^2$ can be compared to pQCD 
when either
$M_V^2$, $t$ or $Q^2$ become larger than the scale $\Lambda$. 
Precisely how the transition from the non-perturbative to the perturbative 
regime is made is currently being determined experimentally.

\section{Photon Dissociation}
\begin{figure}[htb]
\epsfxsize=8.cm
\centering
\leavevmode
\epsfbox[49 147 520 645]{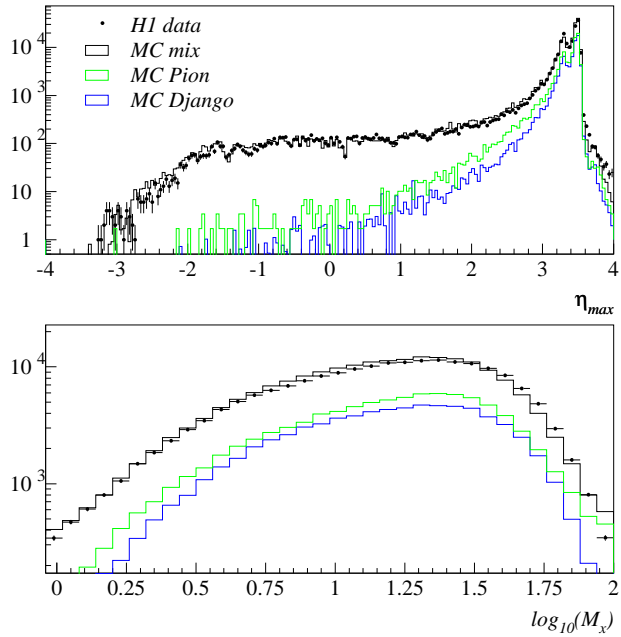}
\caption{Diffractive event selection: H1 analysis of the $\eta_{max}$ 
distribution. 
The upper plot is the $\eta_{max}$ distribution, where a clear excess is seen
over the non-diffractive Monte Carlo's discussed in the text. The lower plot
shows the observed $M_X$ distribution.}
\label{h1c}
\end{figure}

{\it{Diffractive Event Selection:}}
The study of the vector meson resonances enables specific points on 
the $M^2$ scale to be investigated. The inclusive dissociation 
into any low-mass state, $X$, from the (virtual) photon provides additional
information: one of the major advances in the subject of diffraction has
been the observation of large rapidity gap events in DIS and their 
subsequent analysis in terms of a diffractive structure 
function~\cite{Hd1,Zd1}. 
In addition, the relationship between these DIS measurements and those
in the photoproduction regime provide insight into the transition 
of the diffractive structure function in the $Q^2\rightarrow 0$
limit. In these analyses,
the typical signature of diffraction is a rapidity gap,
defined by measuring the maximum pseudorapidity of the 
most-forward going particle
with energy above 400~MeV, $\eta_{max}$, 
and requiring this to be well away from
the outgoing proton direction. A typical requirement of 
$\eta_{max} < 1.5$ corresponds to a low 
mass state measured in the detectors of $\ln(M_X^2) \sim 4$ units 
and a large gap of 
$\ln(W^2) - \ln(M_X^2) \sim 8$ units with respect to the 
outgoing proton (nucleon system).
In order to increase the lever arm in $M_X^2$, 
the H1 analysis has extended the $\eta_{max}$ cuts to 3.2. 
This is achieved by combining the calorimetry information
with the forward muon system and proton remnant taggers.
These extensions enable a cross-section beyond that due to 
simple diffractive processes to be determined at the expense of a 
significant non-diffractive contribution (up to $\simeq 50\%$).
As illustrated in Fig.~\ref{h1c}, 
the Monte Carlo description of the $\eta_{max}$
distributions shows a clear excess over the non-diffractive models 
(labelled MC Django) and a Monte Carlo to describe higher reggeon exchanges
(labelled MC Pion). The data is well-described by a mixture
(labelled MC mix) including an additional contribution due to the pomeron
exchange, both in the
$\eta_{max}$ distribution and the corresponding observed $M_X$ distribution.

\begin{figure}[htb]
\epsfxsize=8.cm
\centering
\leavevmode
\epsfbox[105 240 445 585]{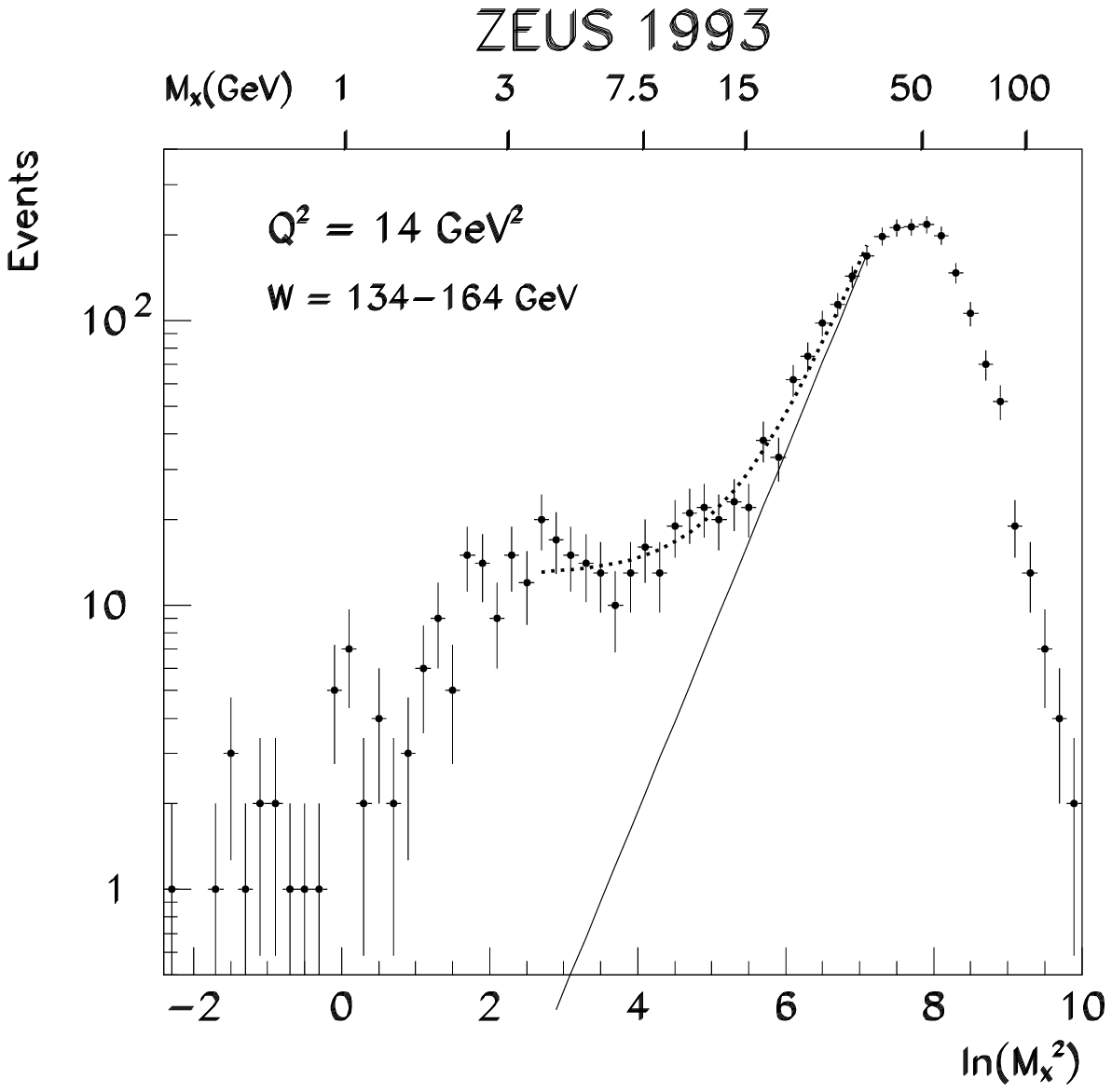}
\caption{Diffractive event selection:
ZEUS analysis of the $\ln M_X^2$ distributions 
for $134<W<164$~GeV and $Q^2 = 14$~GeV$^2$. 
The solid line shows the extrapolation 
of the nondiffractive background determined from the fit to the
data (dotted line) discussed in the text.}
\label{zeusm}
\end{figure}

One of the major uncertainties comes from the estimation of the 
various contributions to the cross-section which depends on 
Monte Carlo techniques. 
This problem has been addressed in a different
way in the ZEUS analysis~\cite{Zd2}.
Here, the mass spectrum, $M_X^2$, 
is measured as a function of $W$ and $Q^2$, 
as shown in Fig.~\ref{zeusm} for a representative interval, where
the measured mass is reconstructed in the calorimeter and 
corrected for energy loss but not for 
detector acceptance, resulting in the turnover at large $M_X^2$. 
The diffractive data are observed as a low mass shoulder at low $M_X^2$. 
which becomes increasingly apparent at higher $W$.
Also shown in the figure are the estimates of the non-diffractive 
contribution 
based on a direct fit to the data, discussed below.

The probability of producing a gap is exponentially suppressed 
as a function of the rapidity gap, and hence as a function of $\ln(M_X^2$),
for non-diffractive interactions. The slope of this exponential
is directly related to the height of the plateau distribution of 
multiplicity in the region of rapidity where the subtraction is made.
The data can thus be fitted to functions of the form
$ dN/d \ln(M_X^2) = D + C {\rm exp}( b \cdot \ln(M_X^2)) $, 
in the region where the detector acceptance is uniform,
where $b$, $C$ and $D$ are determined from the fits.
Here, $D$ represents a first-order estimate of the diffractive contribution
which is approximately flat in $\ln(M_X^2$). 
The parameter which determines the background is $b$.
In general the measured value of $b$ is incompatible with that of the 
ARIADNE Monte Carlo.
This result in itself is interesting, since the fact that ARIADNE approximately
reproduces the observed forward $E_T$ ($\sim$ multiplicity) flow but does not
reproduce the measured value of $b$ suggests that significantly different 
correlations of the multiplicities are present in non-diffractive DIS 
compared to the Monte Carlo expectations. 
This method enables a diffractive cross-section to be determined
directly from the data at the expense of being limited in the range of large 
masses that can analysed. 

\begin{figure}[htb]
\vspace{1cm}
\epsfxsize=8.cm
\centering
\leavevmode
\epsfbox[95 350 485 550]{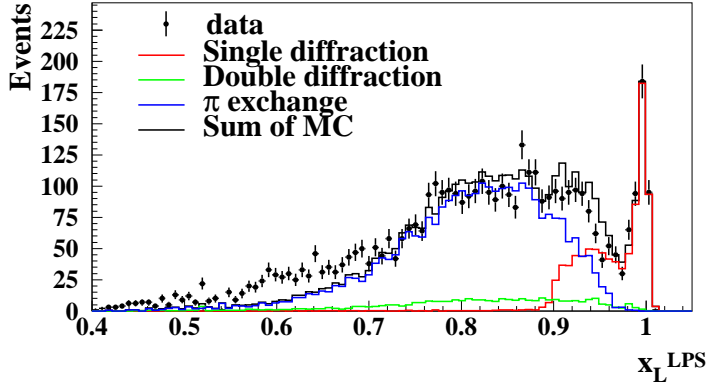}
\caption{Observed $x_L$ spectrum of the ZEUS LPS DIS data. 
The data are described
by a sum of single diffraction (significant at high $x_L$) and pion exchange 
(significant at low $x_L$) with a small contribution due to double
diffraction.}
\label{mxl}
\end{figure}

Finally, 
the advent of the leading proton spectrometers (LPS) at HERA is especially
important in these diffractive measurements, since internal
cross-checks of the measurements as a function of $t$, $M^2$, $W^2$ and $Q^2$
can be performed 
and underlying assumptions can be studied experimentally. Only in these
measurements can we positively identify the diffracted proton and hence 
substantially reduce uncertainties on the non-diffractive and double 
dissociation backgrounds. 
This is illustrated in Fig.~\ref{mxl}, where the $x_L$ (where $x_L = p'/p$) 
distribution includes a clear diffractive peak for $x_L \simeq 1$. 
It should be noted, however, that the contribution from other Reggeon 
exchanges cannot be neglected until $x_L\sgeq 0.99$ (in fact the 
result at lower $x_L$ can be simply interpreted via reggeon 
(approximated by pion) exchange, as discussed below.)
However, new experimental uncertainties are introduced
due to the need for precise
understanding of the beam optics and relative alignment of the detectors.
Reduced statistical precision also results due to the limited 
geometrical acceptance of the detectors ($\simeq$~6\%).

\begin{figure}[htb]
\vspace{1cm}
  \centering
\mbox{
\subfigure[$t$ distribution.]
{\psfig{figure=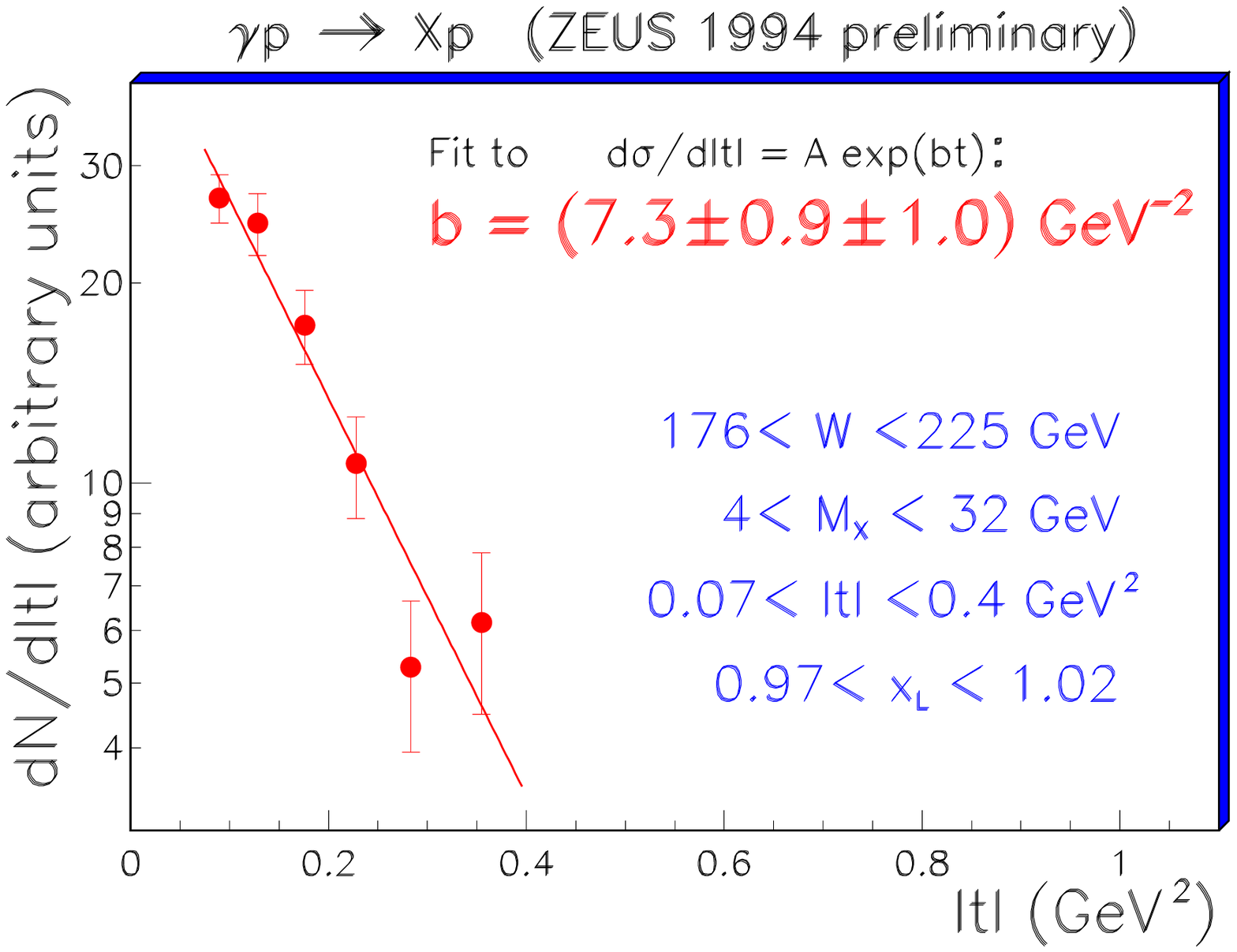,width=.41\textwidth}}\quad
\subfigure[$b$-slopes as a function $W^2$]
{\psfig{figure=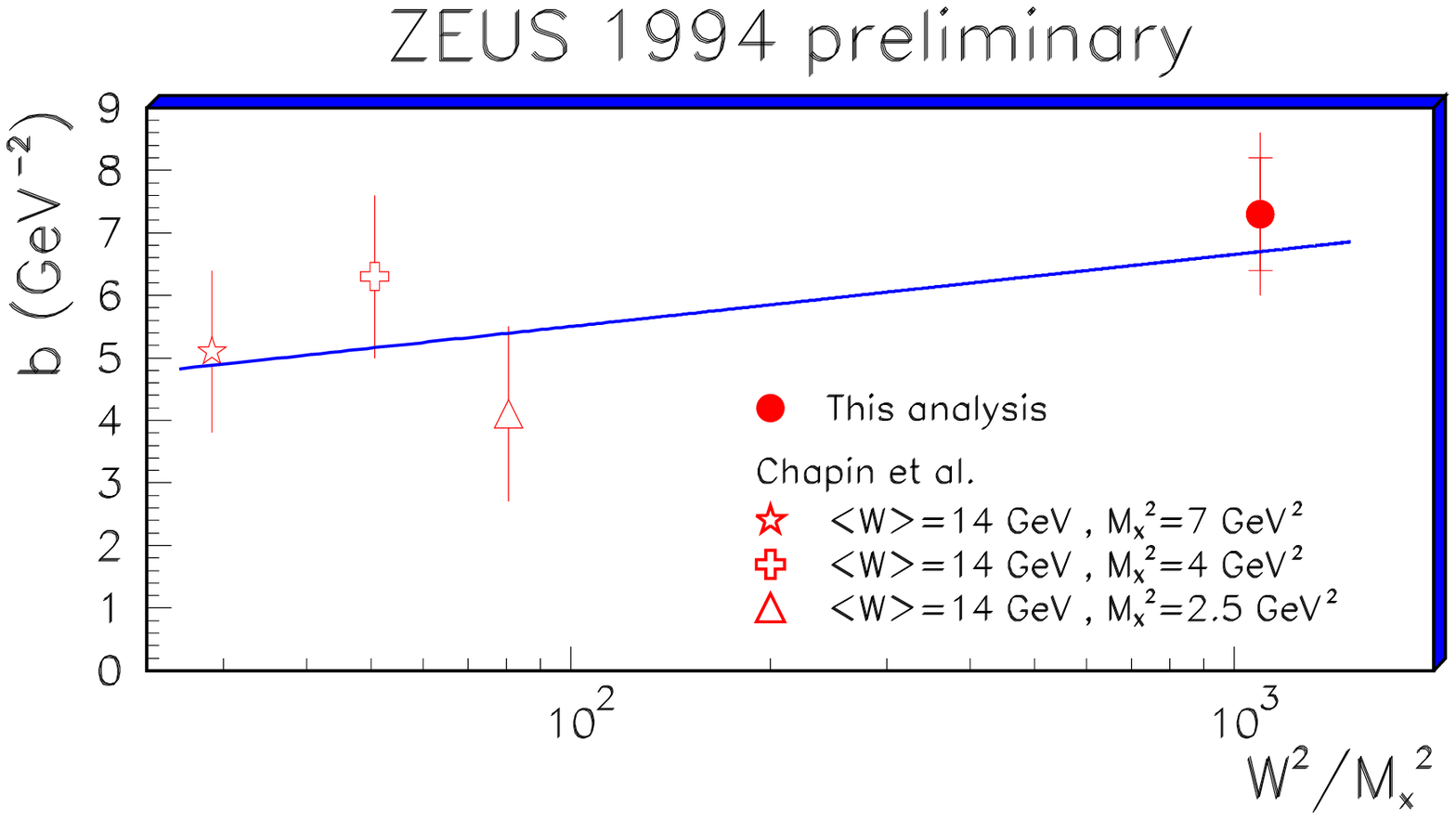,width=.45\textwidth}}
}
  \caption[]{$t$ distribution and 
corresponding $b$-slope as a function of $W^2$ compared to those of
Chapin et al. for inclusive diffractive photoproduction.}
\label{tpho}
\end{figure}

\begin{figure}[htb]
\epsfxsize=10.cm
\centering
\leavevmode
\epsfbox{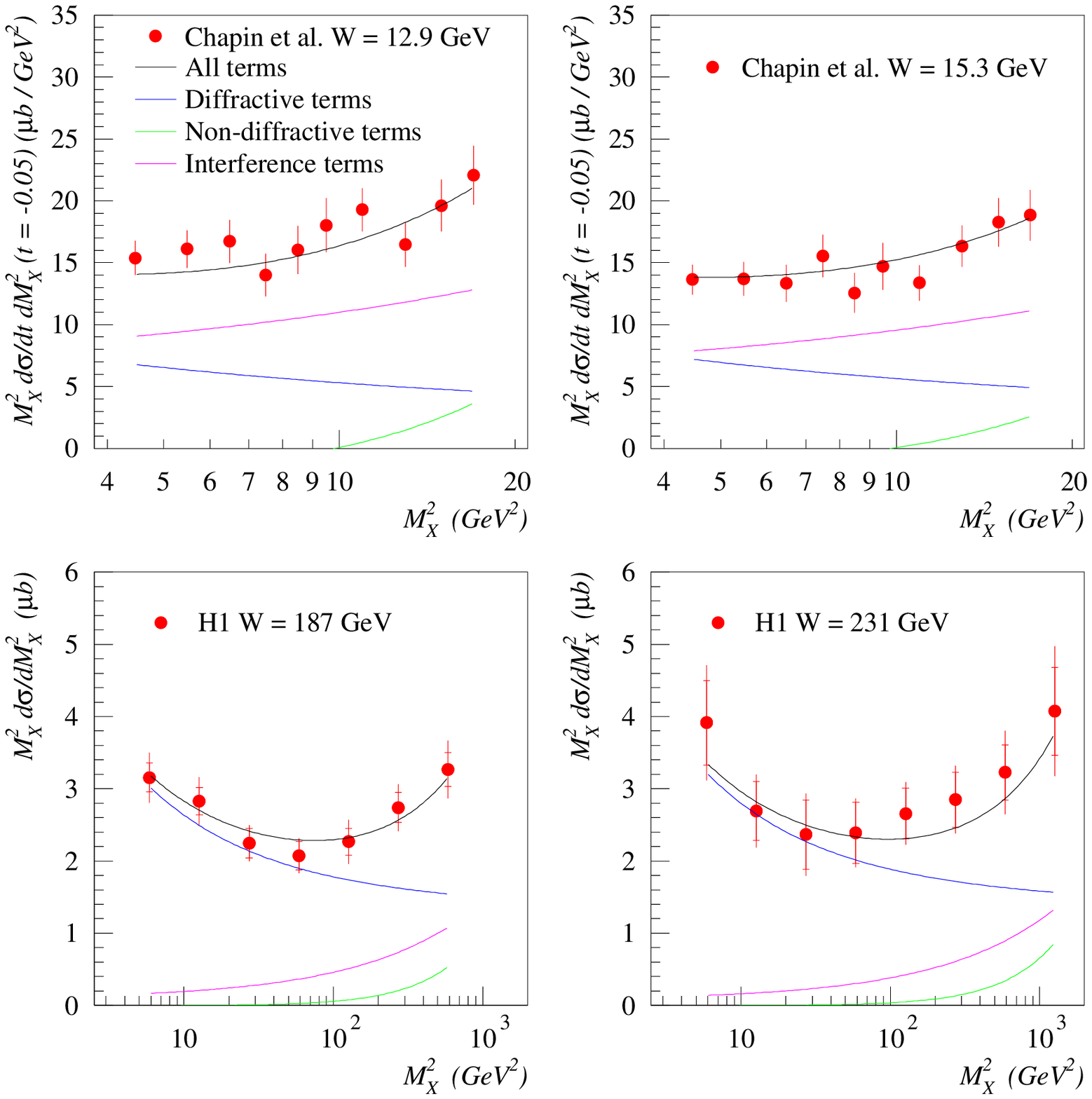}
\caption{Inclusive diffractive photoproduction cross-sections 
of Chapin et al. (lower $W$) and H1 (higher $W$) data
compared to the fit discussed in the text.}
\label{h1phot}
\end{figure}

{\it{Photoproduction Results:}}
ZEUS has measured the photon dissociation $t$ distribution using the LPS, 
as shown in Fig.~\ref{tpho}. 
An exponential fit to the data yields a $b$-slope
parameter, $b= 7.3\pm0.9\pm1.0$~GeV$^{-2}$. A comparison of the data 
with lower $W$ data from Chapin et al. shows that the result is consistent 
with shrinkage, as previously 
discussed in relation to exclusive $\rho^0$ production.
H1 results on the photon dissociation cross-sections as a function of 
$M_X^2$ in two $W$ intervals are shown in Fig.~\ref{h1phot}. 
Regge theory predicts the form of
the cross-section as a function of $M_X$ and $W$, as discussed with respect 
to proton dissociation. The cross-section is therefore fitted to the form 
$$d\sigma/dM_X^2 \propto (1/M_X^2)^{1+\bar{\epsilon}} (W^2)^{\bar{\epsilon}}$$
The contributions due to reggeon exchange are fixed using the 
lower energy data.
Integrating over the $t$ dependence of $\bar{\epsilon}$ yields a value of
$\epsilon = 0.07\pm0.02\pm0.02$(sys)$\pm0.04$(model), again, consistent 
with soft pomeron exchange.

{\it{Deep inelastic structure of diffraction:}}
A new era for diffraction was opened with the study of the dissociation
of $virtual$ photons. 
Here, the photon can be considered as probing the structure of the 
exchanged colourless object mediating the interaction.
The deep inelastic structure of colour singlet exchange is therefore being
studied.
In the presentation of the results, the formalism changes~\cite{ingprytz},
reflecting an assumed underlying partonic description,
and two orthogonal variables are determined 
$$ \xi \equiv \xpom = \frac{(P-P')\cdot q}{P\cdot q} 
\simeq \frac{M_X^2 + Q^2}{W^2 + Q^2}~~~~~~~~~
\beta = \frac{Q^2}{2(P-P')\cdot q} \simeq \frac{Q^2}{M_X^2 + Q^2},$$
where $\xpom$ is the momentum fraction of the pomeron within the proton 
and $\beta$ is 
the momentum fraction of the struck quark within the pomeron.
The structure function is then defined by analogy to that of the 
total $ep$ cross-section
$$
  \frac{d^3\sigma_{diff}}{d\beta dQ^2 d\xpom} = \frac{2 \pi
    \alpha^2}{\beta Q^4} \; Y_+ \;
  F_2^{D(3)}(\beta,Q^2,\xpom),
$$
where 
the contribution of $F_L$ and radiative corrections are neglected
and an integration over the $t$ variable has been performed.

\begin{figure}[htb]
\epsfxsize=10.cm
\centering
\leavevmode
\epsfbox[86 442 479 766]{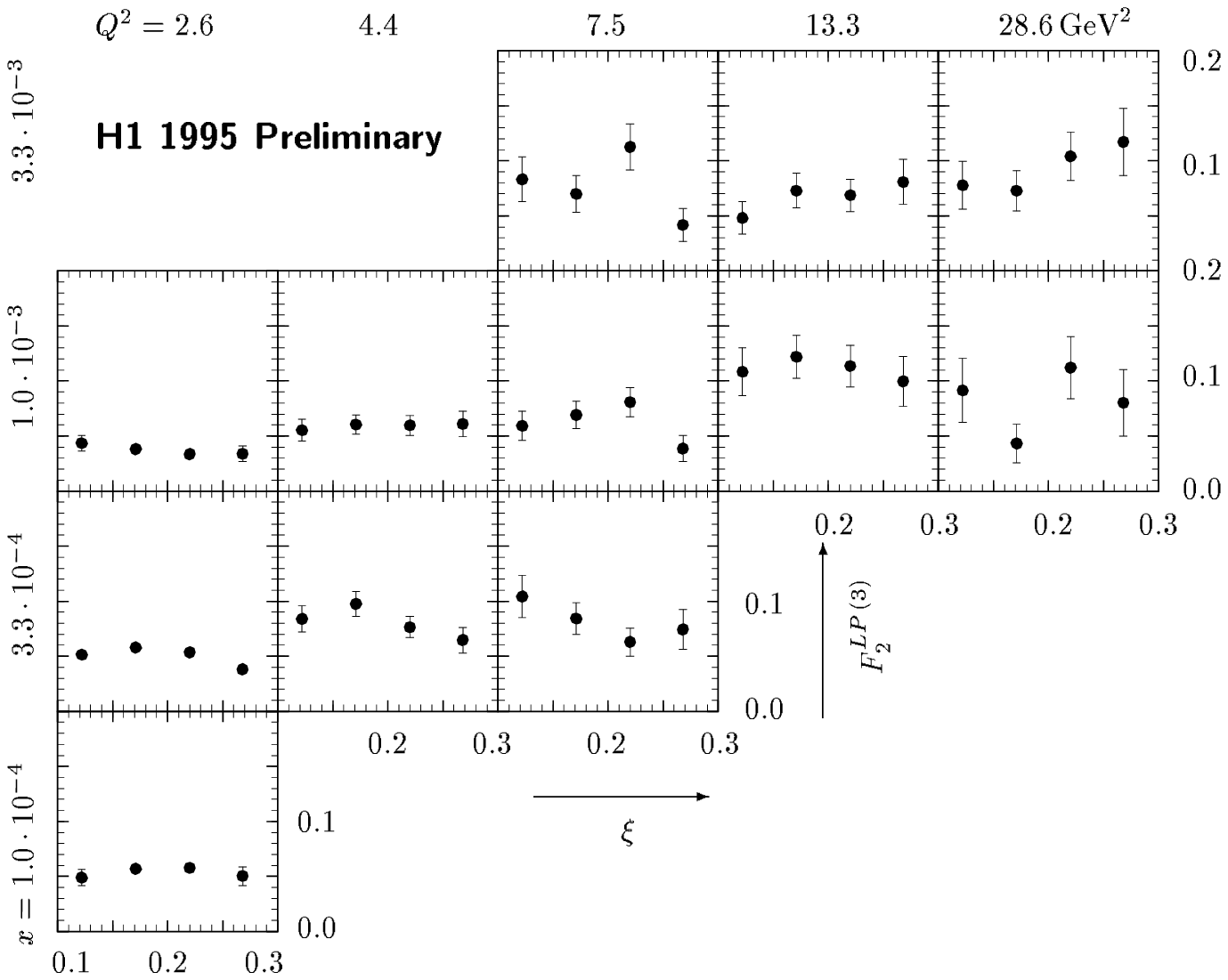}
\caption{H1 data for $F_2^{LP(3)}$ as function of
$\xi$ for the leading proton analysis.}
\label{lp3}
\end{figure}

In addition to the structure of the pomeron, corresponding to large $x_L$,
it is also possible to study the structure of the reggeons that contribute
at lower $x_L$. H1 has analysed the leading proton data at lower $x_L$
($0.7<x_L<0.9$) and employed the formalism noted above to measure the 
structure of the exchange for reasonably forward protons, as
shown in Fig.~\ref{lp3}~\cite{list}.
The data are consistent with a flat $\xi$ dependence in all intervals
of $\beta$ and $Q^2$ i.e. $F_2^{LP(3)} \propto \xi^{n}$.
This is consistent with a factorisable ansatz of
$F_2^{D(3)}(\beta, Q^2, \xi) 
= f_{\regge}(\xi) \cdot F_2^{\regge}(\beta,Q^2), $
where $f_{\regge}(\xi)$ measures the flux of reggeons in the proton
and $F_2^{\regge}(\beta,Q^2)$ is the probed structure of these reggeons.
The exponent of $\xi$ is identified as $n = 1-2\cdot\bar{\eta}$,
where $\bar{\eta}$ measures the effective $\xi$ dependence 
($\equiv W^2$ dependence at fixed $M_X^2$ and $Q^2$) of the cross-section, 
integrated over $t$. 
The data are consistent with $n \simeq 0$, corresponding to
$\bar{\eta} \simeq 0.5$.
These data involve colour singlet exchange and need to be 
explained in terms of QCD, but they are clearly not of a 
diffractive nature.

\begin{figure}[htb]
\epsfxsize=12.cm
\centering
\leavevmode
\epsfbox{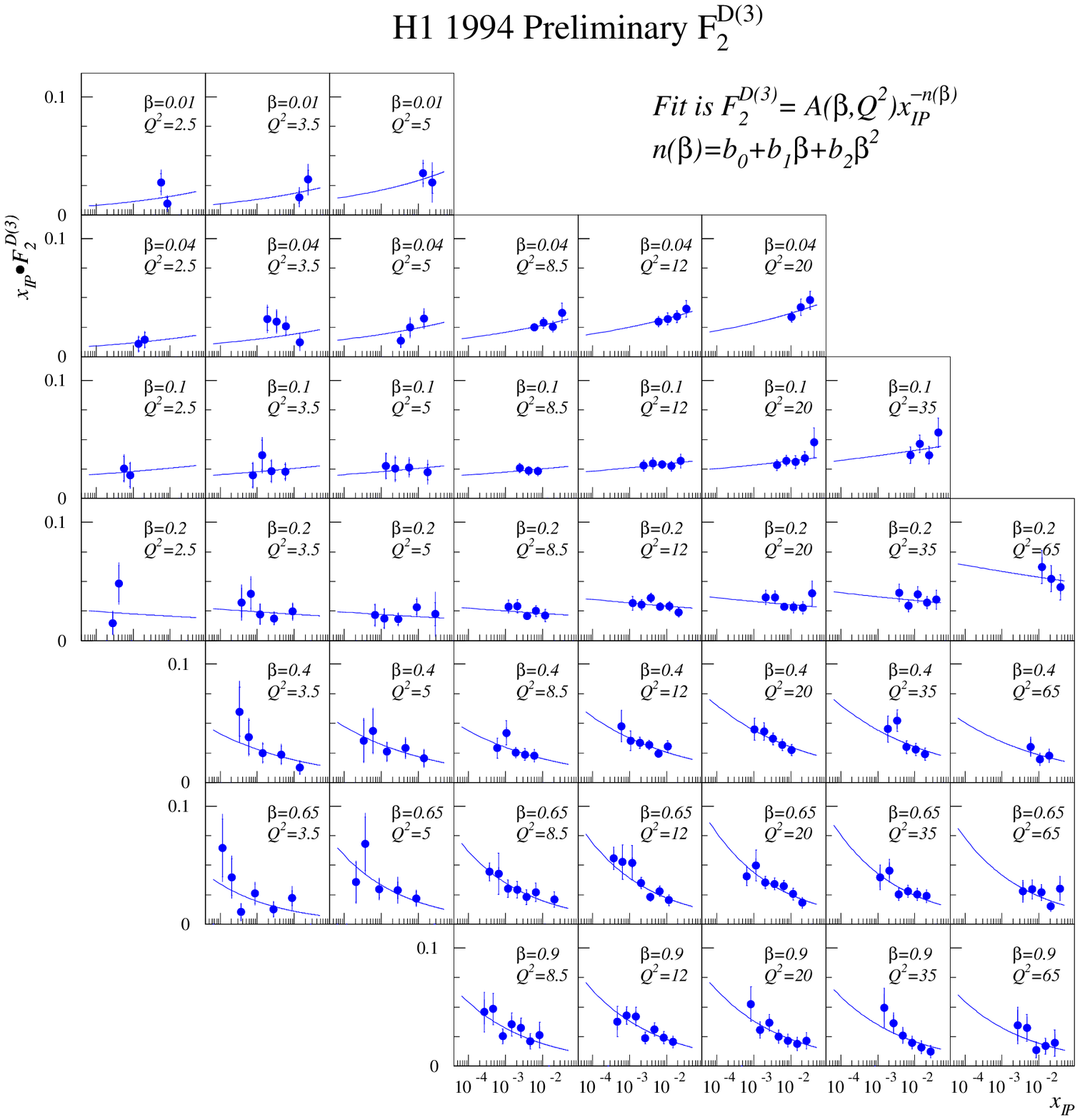}
\caption{H1 preliminary $F_2^{D(3)}$ data 
as function of \xpom in various intervals
of $\beta$ and $Q^2$ ($10^{-4}< \xpom < 0.05; 0.01 < \beta < 0.9; 2.5 < Q^2
< 65$~GeV$^2$). The fitted line corresponds to the QCD fit discussed
in the text.}
\label{h1f2d3}
\end{figure}

The area of interest for diffraction is in the behaviour of small values of 
$\xi \sleq 0.01$, where $\xi$ is now identified as $\xpom$. 
Here, the data fall approximately as
$\xpom^{-1}$ (equivalent to a flat cross-section with increasing $W$) and 
therefore the data are plotted as $\xpom \cdot F_2^{D(3)}$ in 
Fig.~\ref{h1f2d3}.
In the H1 case, the measurement is presented with no explicit 
subtraction for the non-diffractive contribution and quoted for limited 
masses of the dissociated proton system ($M_N < 1.6$~GeV). The measurement
relies upon a good understanding of the various contributions to the 
cross-section in and around the measured region: the control plots in 
Fig.~\ref{h1c} 
illustrate how well this is achieved by combining the different Monte 
Carlo contributions.

\begin{figure}[htb]
\epsfxsize=10.cm
\centering
\leavevmode
\epsfbox{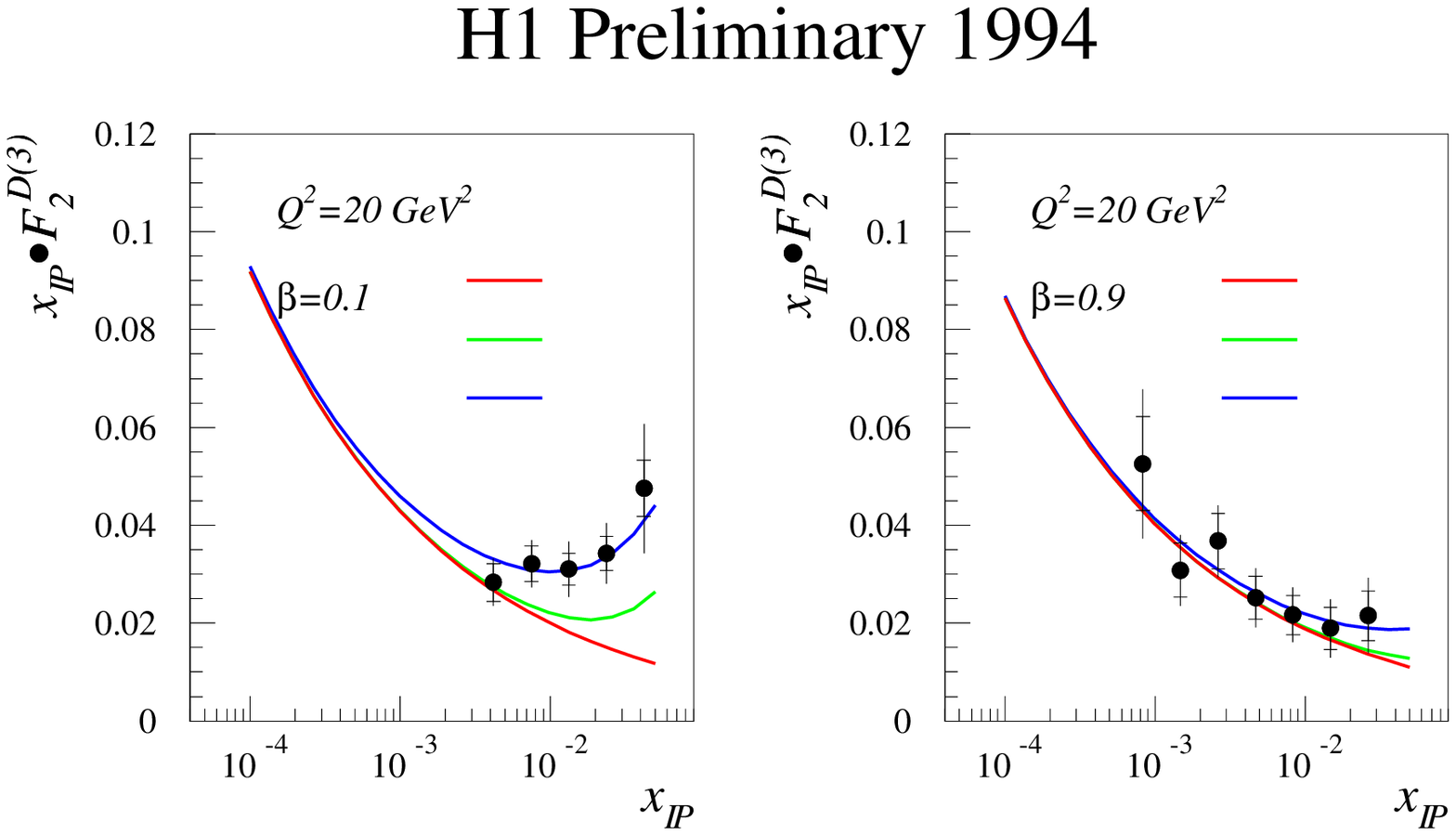}
\caption{Regge fit to sample bins of the H1 $F_2^{D(3)}$ data.}
\label{f2dbits}
\end{figure}

Fits of the form 
$F_2^{D(3)} = A(\beta,Q^2) \cdot \xpom^{n(\beta)}$
are performed 
where the normalisation constants $A(\beta,Q^2)$ are allowed to differ
in each $\beta,Q^2$ interval. 
The fits are motivated by the factorisable ansatz of
$F_2^{D(3)}(\beta, Q^2, \xpom) 
= f_{\pom}(\xpom) \cdot F_2^{\pom}(\beta,Q^2), $
where $f_{\pom}(\xpom)$ measures the flux of pomerons in the proton
and $F_2^{\pom}(\beta,Q^2)$ is the probed structure of the pomeron.
The exponent of \xpom~is identified as $n = 1+2\cdot\bar{\epsilon}$,
where $\bar{\epsilon}$ measures the effective \xpom~dependence 
($\equiv W^2$ dependence at fixed $M_X^2$ and $Q^2$) of the cross-section, 
integrated over $t$, as discussed in relation to exclusive vector meson
production. 
However, the data are now sufficently precise that a single value for
$n$ is not sufficient. In fact, the contribution from reggeon exchange
cannot be excluded since the data extend to $\xpom$ beyond approximately
0.01 and therefore a full fit to the pomeron and reggeon contributions is
required.
A fit to the pomeron and reggeon
contributions has therefore been performed
$$F_2^{D(3)} = \fpom (\beta,Q^2) \cdot (1/\xpom)^{1+2\cdot\bar{\epsilon}}
+ \cregge \fregge (\beta,Q^2) \cdot (1/\xpom)^{1-2\cdot\bar{\eta}}$$
The fit provides a very good overall description of the data with a
$\chi^2$/DoF = 165/156. Here, the function $\fregge$ is taken from the
GRV parametrisation of the pion structure function
and it is assumed that 
the interference between the pomeron and reggeon exchanges is maximal.
This is illustrated for example bins of $\beta$ at fixed $Q^2 = 20$~GeV$^2$
in Fig.~\ref{f2dbits}. 
The lower curve represents the pomeron contribution, the
middle curve corresponds to the sum of the two contributions without taking
into account the (positive) interference and the upper curve is the full fit
result. It should be noted 
that the quark-like reggeon structure function falls like 
$(1-\beta)$ at these $\beta$ values whereas the pomeron structure function 
emerging after integration over $\xpom$ is rather flat: the reggeon
contributions thereforefore play a significant role at smaller values
of $\beta$. Also, for the smallest values of $\beta$, the form of the 
pion (and hence reggeon) structure function is less well known and the
contribution of non-colour singlet exchange contributions starts to become
significant. Assuming $F_L = 0$ and integrating over $t$ with $b =
7$~GeV$^{-2}$ using $\alphappom = 0.25$~GeV$^{-2}$ yields~\cite{Hd2} 
$$\alphapom(0) \equiv 1 + \epsilon = 1.18\pm 0.02 \pm 0.04$$ 
$$\alpharegge(0) \equiv 1 - \eta = 0.6\pm 0.1 \pm 0.3$$ 
 
\begin{figure}[htb]
\epsfxsize=10.cm
\centering
\leavevmode
\epsfbox[60 90 564 753]{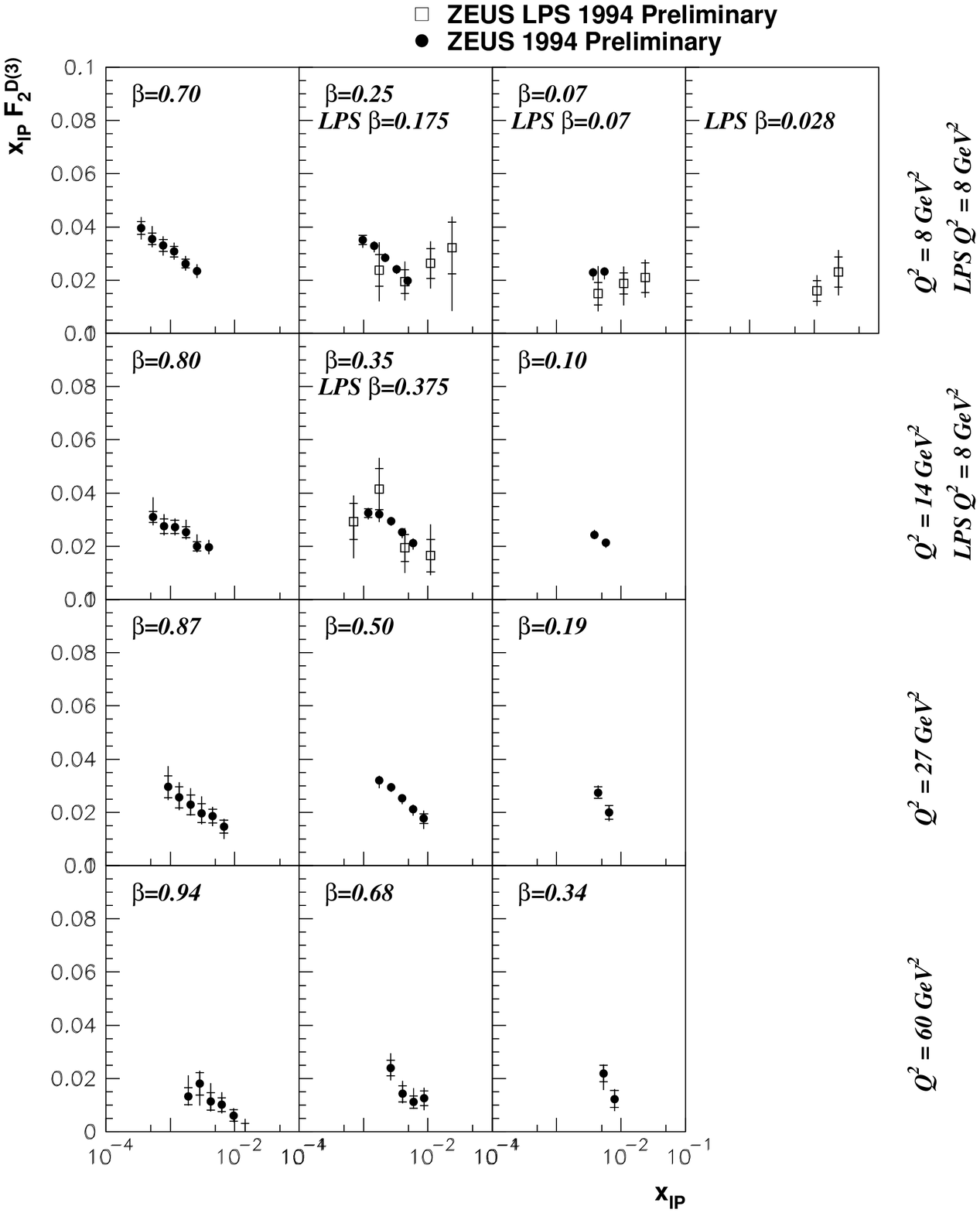}
\caption{Comparison of the ZEUS data for $F_2^{D(3)}$ as function of
\xpom\ for the LPS and $M_X$ analyses in comparable intervals of
$Q^2$ and $\beta$.}
\label{zeusf2d3}
\end{figure}

The ZEUS virtual-photon proton cross-sections 
measured at fixed $M_X^2$ and $W$, measured using the $M_X$ method
(and therefore explicitely subtracting the non-diffractive contribution)
can be converted to $\xpom \cdot F_2^{D(3)}$ at fixed $\beta$ and $\xpom$.
These results are shown in Fig.~\ref{zeusf2d3} 
as the ZEUS($M_X$)~\cite{Zd2} analysis,
compared to the ZEUS LPS analysis of $\xpom \cdot F_2^{D(4)}$,
integrated over the measured $t$ range, 
in comparable intervals of $\beta$ and $Q^2$ as a function of $\xpom$. 

The double dissociation contribution is estimated with 
similar uncertainties to the vector meson case.
Common systematic errors 
are similar to those for the $F_2$ analyses ($\sleq 10\%$) 
with additional acceptance uncertainties due to
variations of the input diffractive Monte Carlo distributions.
The LPS data therefore provide an excellent calibration to cross-check the 
background subtraction methods.
The overall cross-sections in overlapping $\beta$ and $Q^2$ intervals
are in good agreement. In the region of overlap, the H1 and ZEUS LPS 
data points also agree well.
The ZEUS LPS data beyond \xpom of 0.01 again tend to turn over, 
as observed in the H1 data.

\begin{figure}[htb]
\epsfxsize=5.cm
\centering
\leavevmode
\epsfbox[50 415 300 690]{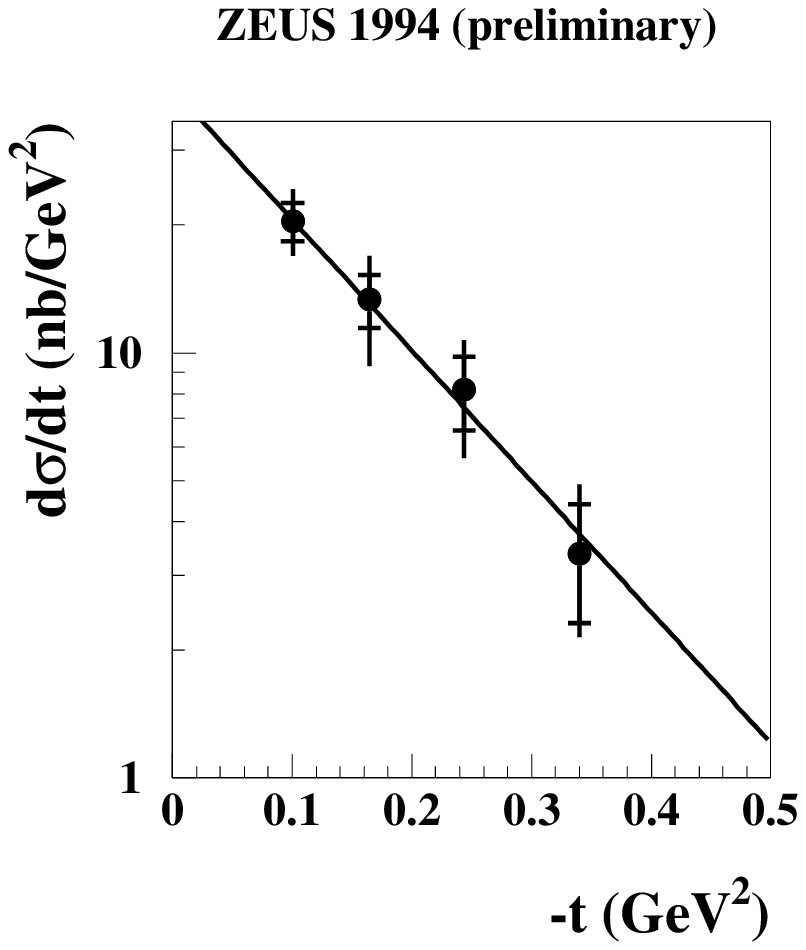}
\caption{ZEUS preliminary $t$ distribution for inclusive diffractive 
DIS data.}
\label{tdis}
\end{figure}

The corresponding ZEUS LPS measurement of the $t$ distribution is shown in
Fig.~\ref{tdis}, measured in the range $x_L > 0.97$,
$5<Q^2<20$~GeV$^2$, $0.015<\beta,0.5$ and $0.073<|t|<0.4$~GeV$^2$.
The slope can be characterised by 
a single exponential fit with 
$b = 7.1 \pm 1.1^{+0.7}_{-1.0}$~GeV$^{-2}$~\cite{Zd2}.
This is very similar to the slope obtained in the photoproduction
case and is somewhat high compared to 
the value of $b\simeq 4.5$ expected for a predominantly hard pomeron
but lies within the range of expectations of $4 \sleq b \sleq 10$.

\begin{figure}[htb]
\epsfxsize=10.cm
\centering
\leavevmode
\epsfbox{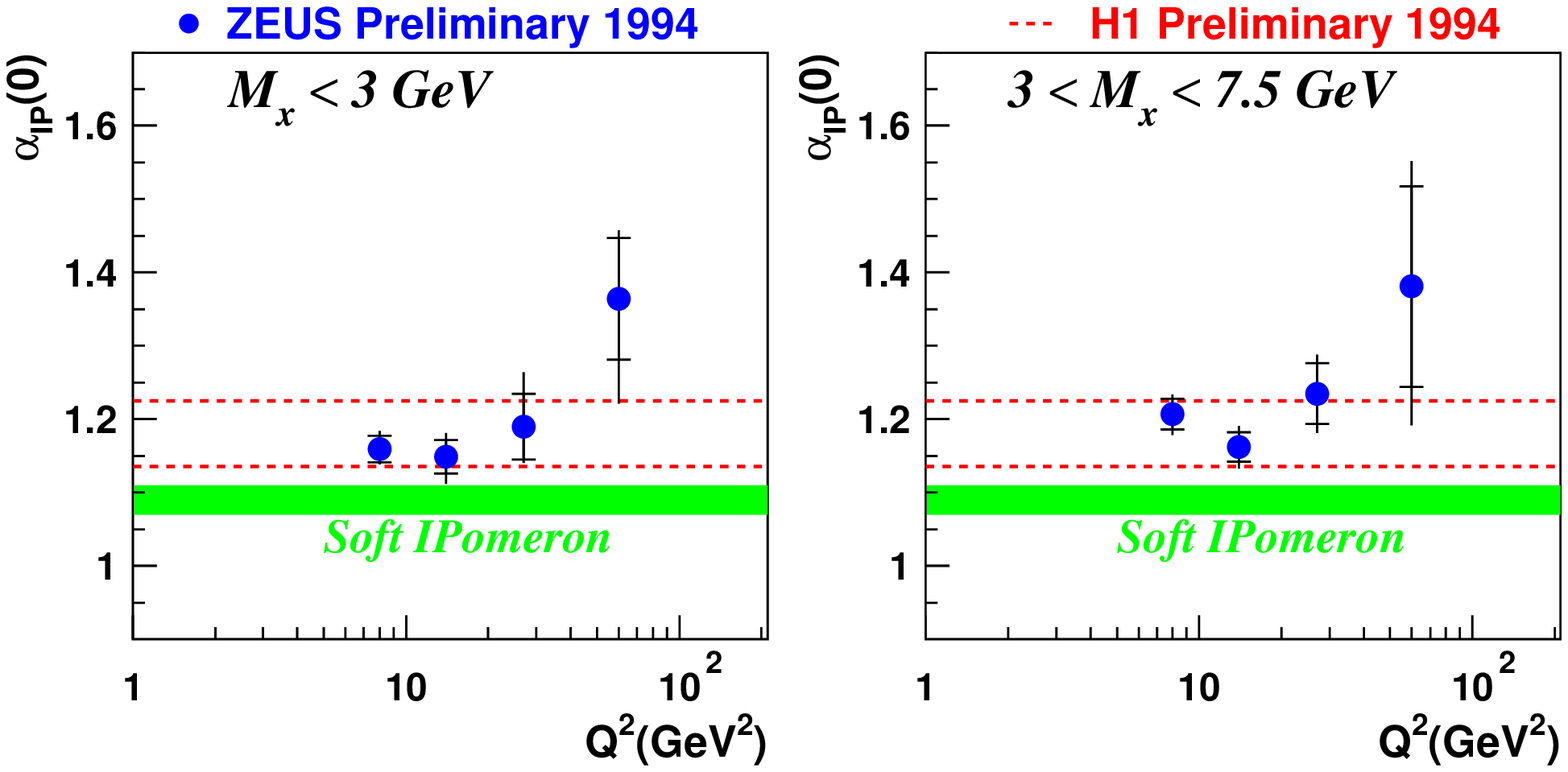}
\caption{$\alphapom(0)$ values extracted from H1 and ZEUS preliminary data. 
The ZEUS data in two $M_X$
intervals as a function of $Q^2$ are compared to the ranges of the 
fit to the H1 data indicated by the dashed lines and the soft pomeron 
prediction indicated by the shaded band.}
\label{alphapom}
\end{figure}

The ZEUS $M_X$ data are largely restricted to the small $\xpom < 0.01$ and
$\beta > 0.1$ region and therefore purely pomeron contributions are 
significant.
A direct fit to the data in fixed $M_X$ intervals yields a similar value
for $\alphapom(0)$. This is illustrated in Fig.~\ref{alphapom} where the 
the $Q^2$ dependence of the extracted $\alphapom(0)$ is compared to the H1
result as well as the soft pomeron prediction.
The results for $\alphapom(0)$, obtained using different experimental
methods, are compatible. 
These values are however incompatible 
with the predicted soft pomeron behaviour
of $\alphapom(0) = 1.08 \pm 0.02$.
As noted previously in relation to exclusive $\rho^0$ production
(corresponding to the large $\beta$ region), the
contribution of longitudinal photons is significant.
An upper estimate of the effect of $\sigma_L$ made by 
assuming $\sigma_L = (Q^2/M_X^2) \sigma_T$ rather than $\sigma_L = 0$
increases the measured value of $\alphapom(0)$ by about 0.05.
Similarly, if no shrinkage were assumed, the 
value of $\alphapom(0)$ would increase by $\alpha'\cdot 1/b = 0.035$.

The values can be compared with $\bar{\epsilon} \simeq$ 0.2
obtained from the exclusive photoproduction of $J/\psi$ mesons and the 
ZEUS exclusive $\rho^0$ electroproduction data at large $Q^2$.
These values are also compatible with fits to the inclusive DIS $F_2$ 
(i.e. the dependence of the corresponding total cross-sections)
which yield $\epsilon \simeq$ 0.2 to 0.25
in the measured $Q^2$ range~\cite{levy}.
In the model of Buchm\"uller and Hebecker~\cite{buch}, 
the effective exchange is 
dominated by one of the two gluons. In terms
of $\epsilon$, where the optical theorem is no longer relevant,
the diffractive cross-section
would therefore rise with an effective 
power which is halved to $\epsilon \simeq$ 0.1 
to 0.125. Given the uncertainties, 
the measured values are within the range of 
these estimates.

\begin{figure}[htb]
\vspace{0.5cm}
\epsfxsize=8.cm
\centering
\leavevmode
\epsfbox{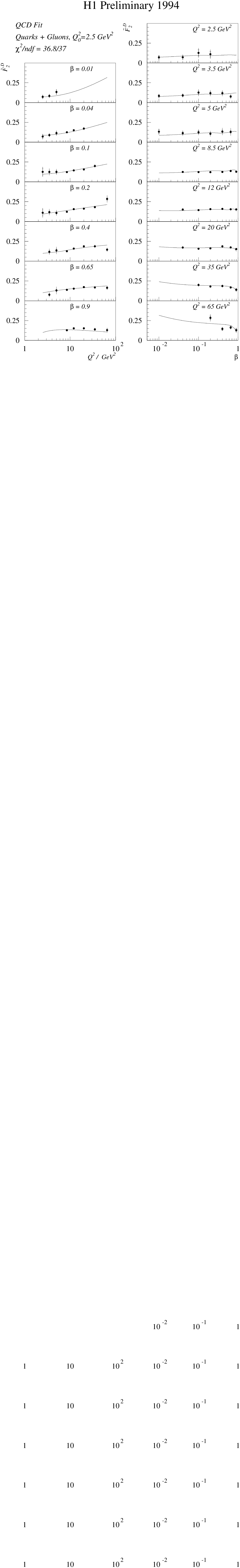}
\vspace{0.5cm}
\caption{H1 preliminary data on $\tilde{F}_2^D$($\beta, Q^2$) as a function
of $Q^2$ ($\beta$) at fixed $\beta$ ($Q^2$). The data are compared to
preliminary next-to-leading-order QCD fits where gluons also contribute at 
the starting scale $Q_0^2 = 2.5$~GeV$^2$, resulting in a fit where gluons 
carry $\sim$ 80\%
of the momentum, indicated by the full line ($\chi^2$/DoF = 36.8/37).
}
\label{h1f2dt}
\end{figure}

The overall cross-sections in each $\beta$, $Q^2$ interval are similar
and one can integrate over the measured \xpom~dependence
in order to determine $\tilde{F}_2^D$($\beta, Q^2$), a quantity which measures
the internal structure of the pomeron up to an arbitrary integration 
constant. 
In Fig.~\ref{h1f2dt} the H1 data are
compared to preliminary QCD fits~\cite{Hd2}. 
The data are fitted in the range $0.0003 < \xpom < 0.05$ using the functional 
form shown in Fig.~\ref{h1f2d3}.

\begin{figure}[htb]
\centering
\leavevmode
\epsfig{file=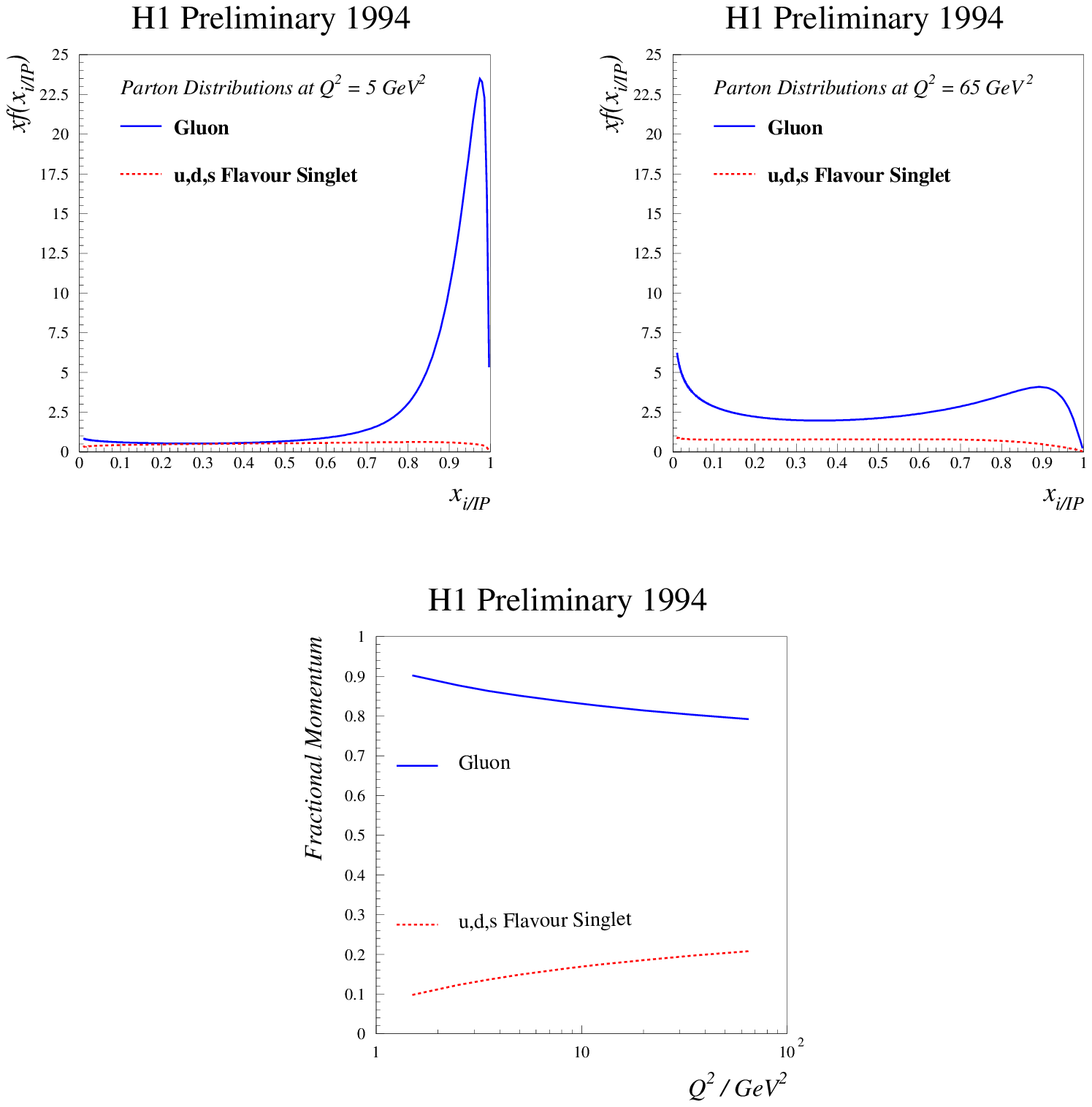,angle=270,height=10cm}
\caption{H1 preliminary parton distributions of the pomeron.
The upper plots show the momentum distributions 
at low $Q^2 = 5$~GeV$^2$ and high $Q^2 = 65$~GeV$^2$
as a function of $z\equiv \xipom$.
The lower plot indicates the 
fraction of the total momentum carried by gluons and quarks
as a function of $Q^2$.
}
\label{h1pdfs}
\end{figure}

The general
conclusions from the $\beta$ dependence are that the pomeron has a 
rather flat structure as a function of $\beta$.
This is typically characterised by a 
symmetric $\beta(1-\beta)$ dependence, but an additional
significant contribution at low $\beta$ is required
which has been fitted in the ZEUS analysis~\cite{Zd1}. 
The $Q^2$ behaviour is broadly scaling, consistent with a partonic
structure of the pomeron. Probing more deeply, however, a characteristic 
logarithmic rise of $\tilde{F}_2^D$ is observed in all $\beta$ intervals.
Most significantly, at large $\beta$ a predominantly quark-like object
would radiate gluons resulting in negative scaling violations as in the 
case of the large-$x$ ($\sgeq 0.15$) behaviour of the proton.
The question of whether the pomeron is predominantly quarks or gluons,
corresponding to a ``quarkball" or a ``gluemoron"~\cite{cf},
has been tested quantitatively by H1 using QCD fits to 
$\tilde{F}_2^D$~\cite{Hd2}. A flavour singlet 
quark density input of the form $zq(z) = A_q \cdot z^{B_q}(1-z)^{C_q}$, where
$z$ is the momentum fraction carried by the quark, yields a high 
$\chi^2$/DoF~=~95/39, since the characteristic $Q^2$ behaviour is not
reproduced. 
Adding a gluon contribution of similar form gives an excellent 
description of the data with $\chi^2$/DoF~=~36.8/37,
as shown in Fig.~\ref{h1f2dt}. 
In Fig.~\ref{h1pdfs}, the corresponding parton distributions 
are shown. 
In general, the fits tend to favour inputs where 
the gluon carries a significant fraction, $\sim$ 70 to 90\%,
of the pomeron's momentum.

\section{Hadronic final states and jet structure}
The measurements of the scaling violations of the structure function of
the pomeron provide a method to determine the parton distributions 
of the pomeron. The question of whether such an approach is useful can 
be addressed by applying these parton distributions to calculations
for other processes which are directly sensitive to this partonic 
structure.

\begin{figure}[htb]
  \centering
\mbox{
\subfigure[$<p_T^{*2}>$ versus $x_F$.]
{\psfig{figure=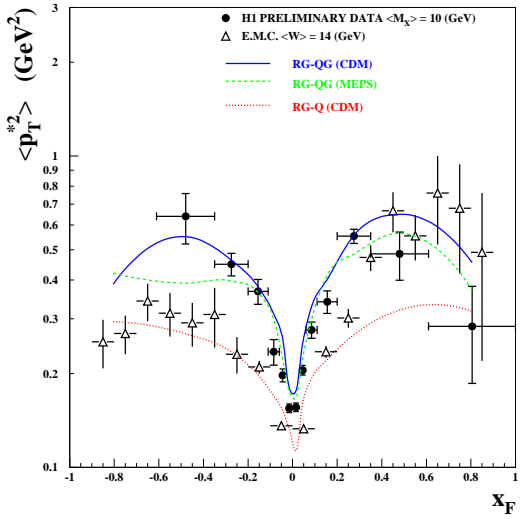,width=.45\textwidth}}\quad
\subfigure[$<{\rm thrust}>$ versus 1/$M_X$.] 
{\psfig{figure=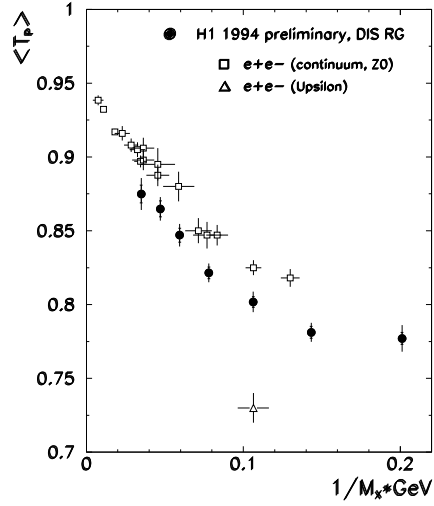,width=.45\textwidth}}
}
  \caption[]{H1 preliminary hadronic final state distributions. 
(a) $<p_T^{*2}>$ versus $x_F$ compared to EMC inclusive DIS data at similar 
$W$ values and the RAPGAP Monte Carlo predictions discussed in the text. 
(b) $<{\rm thrust}>$ versus 1/$M_X$ compared to $e^+e^-$ data at similar 
1/$W$ values.}
\label{seathr}
\end{figure}

Historically, the measurements of $<p_T^{*2}>$, the mean transverse 
momentum-squared of the outgoing hadrons, as a function of 
$x_F = p_L/p_L^{max}$, the scaled longitudinal momentum distribution,
provided insight into the structure of the proton. Here, 
the variables are measured in the hadronic centre of mass frame 
and with respect to the virtual photon-proton 
axis which is equivalent to the virtual photon-pomeron axis for
small values of $t$. 
In Fig.~\ref{seathr}(a), the H1 $\gamma^*\pom$ data (full circles)
are compared to the EMC $\gamma^*p$ data at similar 
$M_X\equiv W$ values. The data are also compared to the RAPGAP (RG)
Monte Carlo predictions incorporating quarks and gluons (-QG) and 
quarks only (-Q)~\cite{chris}. (MEPS) and (CDM) refer to the Matrix Elements plus 
Parton Showers and Colour Dipole Model fragmentation schemes, respectively.
The H1 data
are approximately symmetric about $x_F = 0$ with a 
relatively large $<p_T^{*2}>$ peaking around 0.6 GeV$^2$.
The symmetry and relatively large $p_T^*$  values reflect the 
underlying boson-gluon fusion process where a ``leading" gluon
from the pomeron interacts with the virtual photon.
This behaviour is in contrast to the EMC $\gamma^*p$ data where 
QCD radiation is suppresed in the negative-$x_F$ (proton remnant) region.
Quantitatively the RAPGAP Monte Carlo which incorporates the pomeron 
parton densities (-QG) gives a good description of the data, provided that 
quarks and gluons are incorporated whereas a model with only quark (-Q)
fails to describe the data. These conclusions are relatively
independent of the fragmentation scheme, but the colour dipole model tends to
give a better description of the data.

Similarly, event shape variables have been used at $e^+e^-$ colliders
in order to establish the existence of gluon Bremsstrahlung radiation.
In this case, the measurement of e.g. mean thrust (the mean value of
the scaled longitudinal momentum with respect to the axis which 
maximises this value) is sensitive to the gluon-induced diagrams.
A comparison of $<$thrust$>$ 
with $e^+e^-$ annihilation experiments as a function of
the reciprocal of hadronic centre of mass is shown in Fig.~\ref{seathr}(b).
The diffractive data exhibit lower thrust values compared to $e^+e^-$ 
data for all values of $M_X$. This additional
broadening is due to the boson gluon fusion process which has no analogue 
in $e^+e^-$ annihilation continuum region.

\begin{figure}[htb]
  \centering
\mbox{
\subfigure[$\beta^{OBS}$ distribution.]
{\psfig{figure=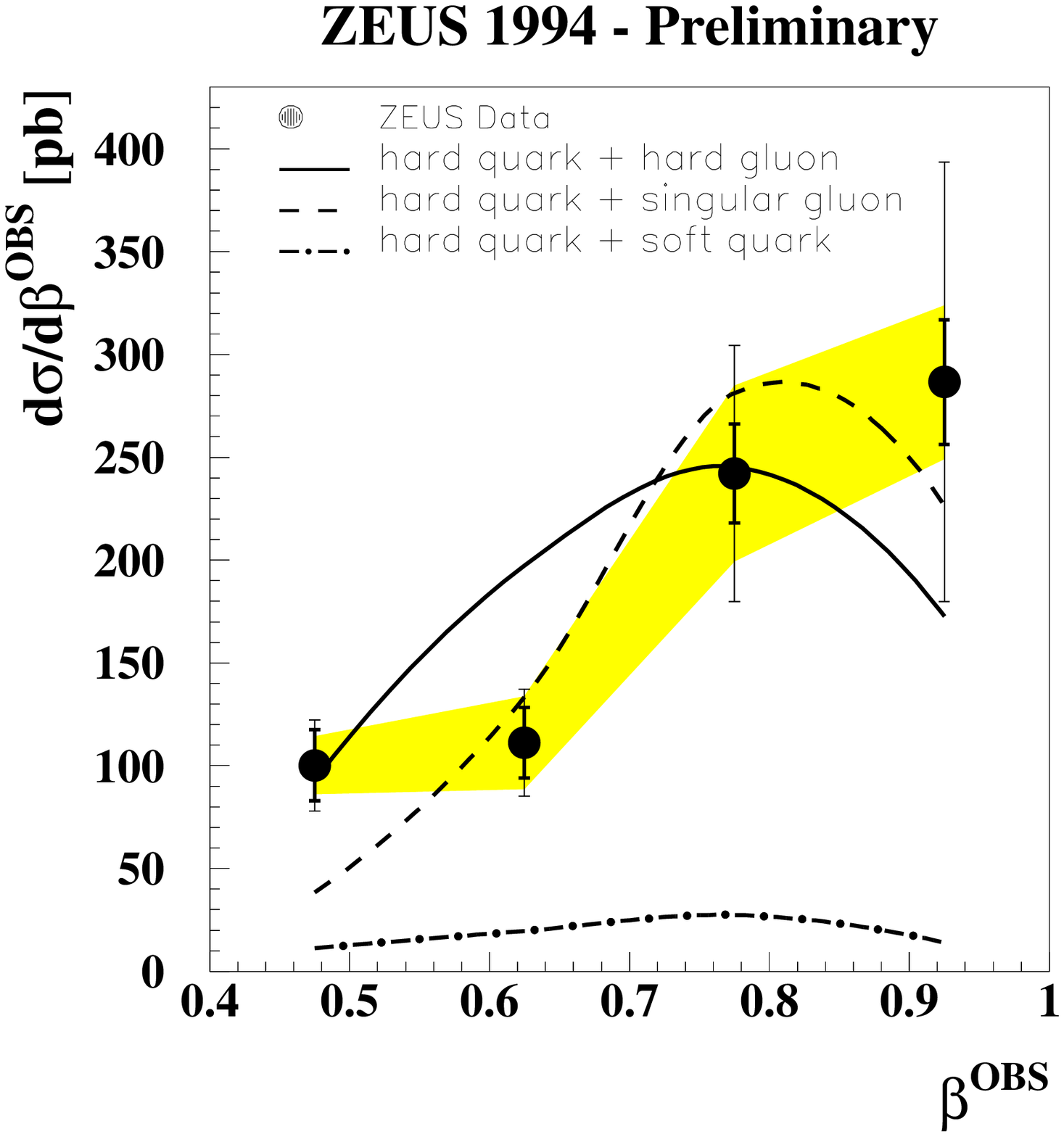,width=.45\textwidth}}\quad
\subfigure[$x_\gamma^{OBS}$ distribution.] 
{\psfig{figure=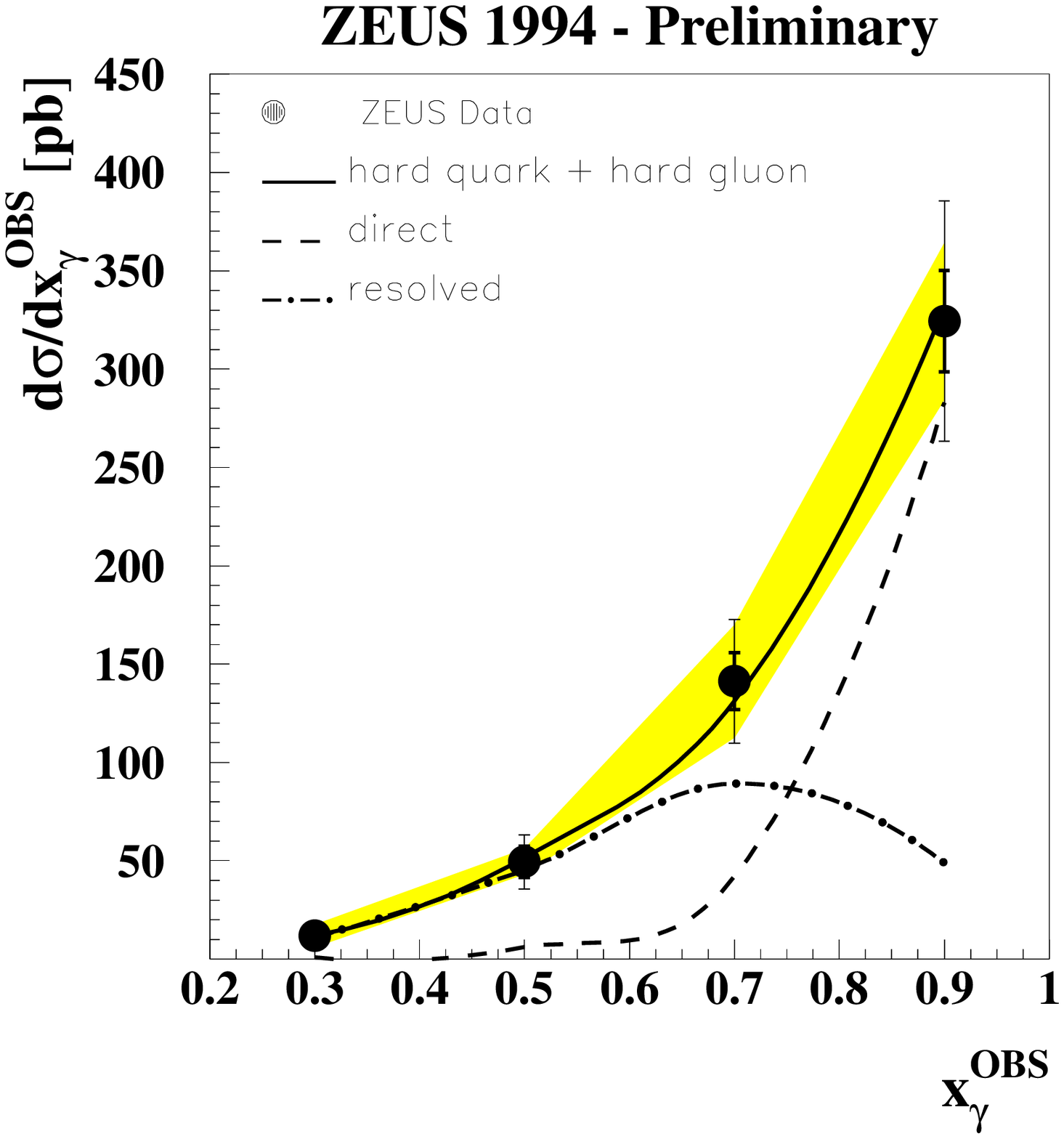,width=.45\textwidth}}
}
  \caption[]{ZEUS preliminary dijet cross-sections from large 
$E_T^{jet}$ photoproduction data with a large rapidity gap for (a) the pomeron and
(b) the photon. The shaded band                    
represents the (correlated) energy scale uncertainty. The data are 
compared to various combinations of quark and gluon input distributions of 
the pomeron for the QCD fits discussed in the text.} 
\label{betaxg}
\end{figure}

The general increase in thrust with increasing $M_X$ (decreasing $1/M_X$)
is indicative of jet production. 
The question of the constituent content of the pomeron can also be addressed
via measurements of diffractively produced jets in the photoproduction
data~\cite{Zjet}. Jets are reconstructed at large $W$ ($134< W < 277$~GeV) 
using the cone algorithm with unit cone radius and two jets with 
$E_T^{jet} > 6$~GeV.
The diffractive
contribution is identified as a tail in the $\eta_{max}$ distribution 
of these events above the 
PYTHIA~5.7~\cite{pythia} Monte Carlo expectation. 
In Fig.~\ref{betaxg} the measured
cross-section is compared to various model predictions as a function of
$\beta^{OBS}$, an estimator of the fraction of the pomeron momentum transferred
to the dijet system. 

The non-diffractive contribution estimated from
PYTHIA (not shown) is significantly lower than the data.
Here, standard photon and proton parton distributions are adopted and
the overall scale, which agrees with the non-diffractive data normalisation,
is set by $E_T^{jet}$. Also shown are the predicted diffractive cross-sections
from the LO QCD calculation plus parton showers of POMPYT, using a 
hard ($z(1-z)$) quark combined with either a  
hard, soft (($1-z)^5$) or singular gluon 
where a Donnachie-Landshoff flux factor is adopted.
Sampling low-energy (soft) gluons corresponds to a small cross-section
and can be discounted, 
whereas high-energy (hard) gluons and/or quarks can account for the 
cross-section by changing the relative weights of each contribution.
The shape of the $\beta^{OBS}$ distribution is clearly sensitive to the 
shape of the input gluon distribution.

The $x_\gamma^{OBS}$ distribution for these events, where $x_\gamma^{OBS}$ 
is the 
corresponding estimator of the fraction of the photon momentum transferred to
the dijet system, 
is peaked around 1, indicating that at these $E_T^{jet}$ values
a significant fraction of events is due to direct processes where the 
whole photon interacts with the pomeron constituents. 

\begin{figure}[htb]
\epsfxsize=5.cm
\epsfysize=5.cm
\centering
\leavevmode
\epsfbox{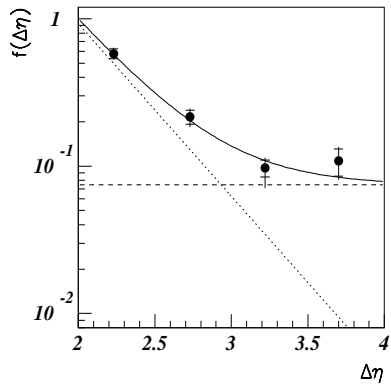}
\caption{ZEUS gap-fraction, $f(\Delta\eta)$, as a function of the rapidity gap
between the two jets compared with the result of a fit to an exponential plus 
a constant.}
\label{zeuscs}
\end{figure}

So far we have only considered the case of small-$t$ diffraction with respect
to the outgoing proton. Further insight into the diffractive exchange process
can be obtained by measurements of the rapidity gap between jets. Here, 
a class of events is observed with little hadronic
activity between the jets~\cite{Zt}. 
The jets have $E_T^{jet} > 6$~GeV and are separated by a pseudorapidity 
interval ($\Delta\eta$) of up to 4 units.
The scale of the momentum transfer, $t$, is not precisely defined but 
is of order $(E_T^{jet})^2$.
A gap is defined as the absence of particles with
transverse energy greater than 300~MeV between the jets.
The fraction of events containing a gap is then measured as a function of
$\Delta\eta$, as shown in Fig.~\ref{zeuscs}.  
The fit indicates the sum of an exponential behaviour, as
expected for non-diffractive processes and discussed in relation to the 
diffractive DIS data, and a flat distribution expected for diffractive
processes. At 
values of $\Delta\eta \sgeq 3$, an excess is seen with a constant fraction
over the expectation for non-diffractive exchange 
at $\simeq 0.07\pm 0.03$.
This can be interpreted as evidence for large-$t$ diffractive scattering.
In fact, secondary interactions of the photon and proton remnant
jets could fill in the gap and therefore the underlying process could play
a more significant r\^ole. 
The size of this fraction is relatively large when compared to a similar
analysis by D\O~and CDF where a constant fraction at $\simeq 
0.01$ is observed~\cite{D0,CDF3}, as discussed below.

\section{Diffraction at the Tevatron}
The study of diffraction benefits by looking at information from other
types of interaction. In $e^+e^-$ where there is no complex colour state in 
the interacting beams, various searches for rapidity gaps yield results
which can be explained in terms of exponentially-suppressed colour exchange
and these data do not require the introduction of a ``pomeron"~\cite{SLD}. 
In contrast,
the earliest work on diffractive processes focussed on $pp$ collisions where
two complex colour states interact. The latest studies at the Tevatron
complement those at HERA and allow tests of factorisation in $ep$ compared 
to $\bar{p}p$. 

\begin{figure}[htb]
\epsfxsize=12.cm
\centering
\leavevmode
\epsfbox[100 550 440 715]{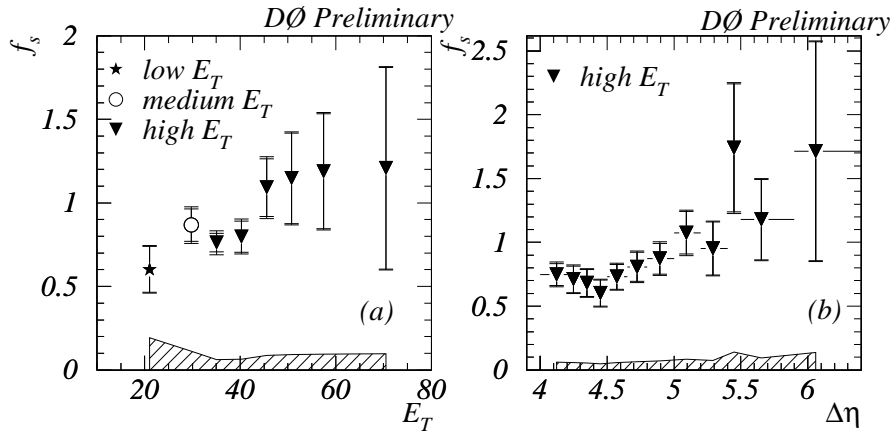}
\caption{Percentage gap-fraction, $f_s$, as a function of (a) the average
dijet $E_T^{jet}$ and 
(b) the rapidity gap between the two jets for the high $E_T^{jet}$ jet sample.}
\label{d0cs}
\end{figure}

The studies of colour singlet exchange between jets at HERA was inspired by
earlier studies at the Tevatron. These studies determined percentage 
gap-fractions of $1.07\pm 0.10^{+0.25}_{-0.13}$\% for D\O~\cite{D0} 
and $0.86\pm0.12$\% for CDF~\cite{CDF3}. 
The behaviour of this colour singlet fraction as a function of average 
dijet energy is shown in Fig.~\ref{d0cs}(a)
for jet $E_T^{jet}$ thresholds of 15~GeV~(low), 25~GeV~(medium) and 
30~GeV~(high)~\cite{perkins}.
In Fig.~\ref{d0cs}(b), the gap-fraction is examined as a function of 
$\Delta\eta$. A simple two-gluon exchange model with no additional QCD
dynamics would tend to produce a flat gap-fraction, but
the tendency towards an increasing gap-fraction 
with $\Delta\eta$ indicates an additional dynamical mechanism
is necessary to describe the data.  
  
The differences in the overall gap-fractions observed at HERA near 10\%,
compared to those at Fermilab of approximately 1\%, may reflect 
the higher 
$W$ values of the Tevatron compared to HERA. But the fact that this 
difference is so large indicates differences in the underlying 
high-$x_\gamma$ $\gamma p$ interactions compared to the relatively low-$x$ 
$p\bar{p}$ interactions where spectator interactions are more likely to fill 
in the gap.

\begin{figure}[htb]
  \centering
\mbox{
\subfigure[D\O~dijet signal distribution.]
{\psfig{figure=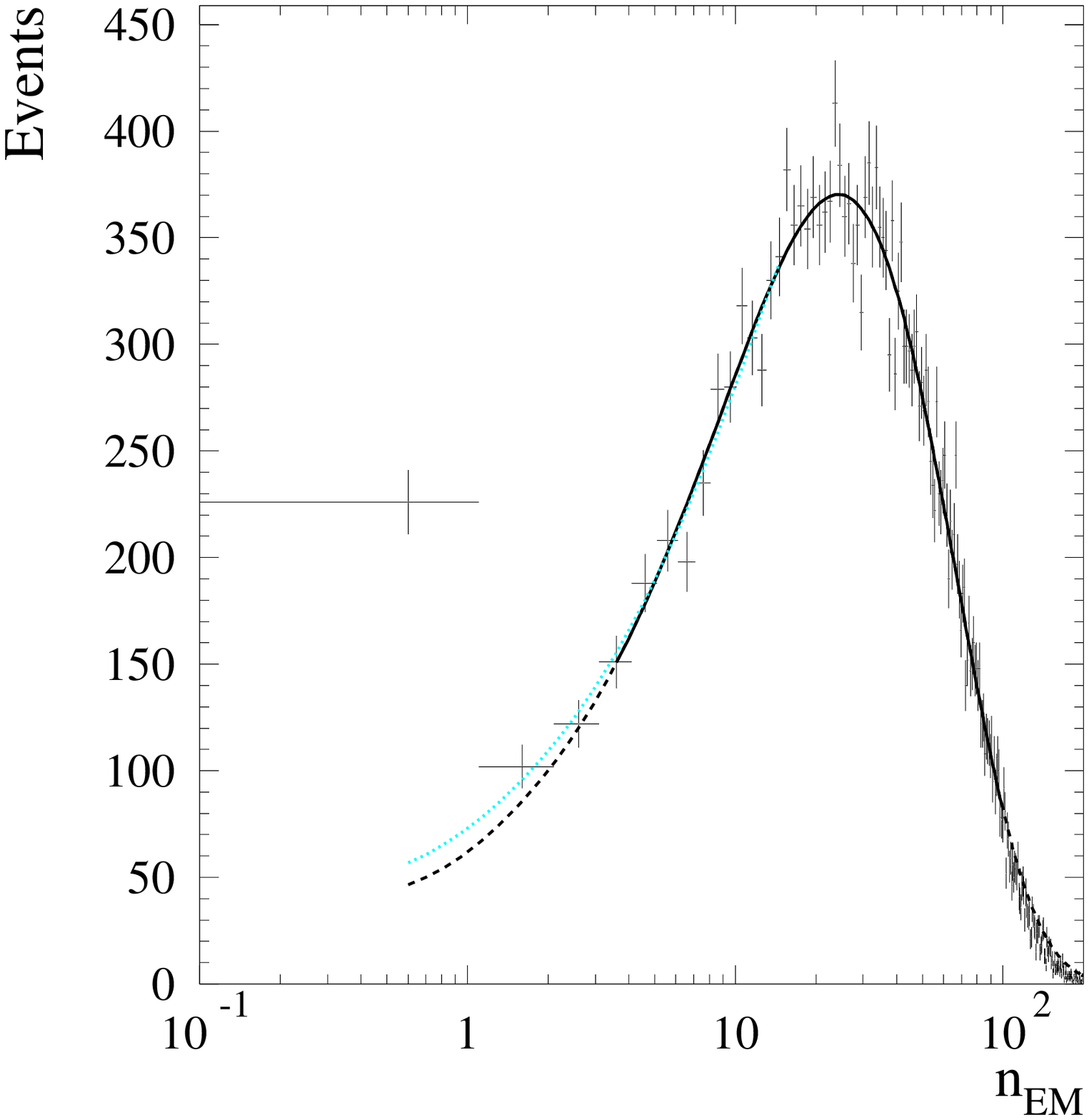,width=.4\textwidth}}\quad
\subfigure[CDF dijet signal distribution.] 
{\psfig{figure=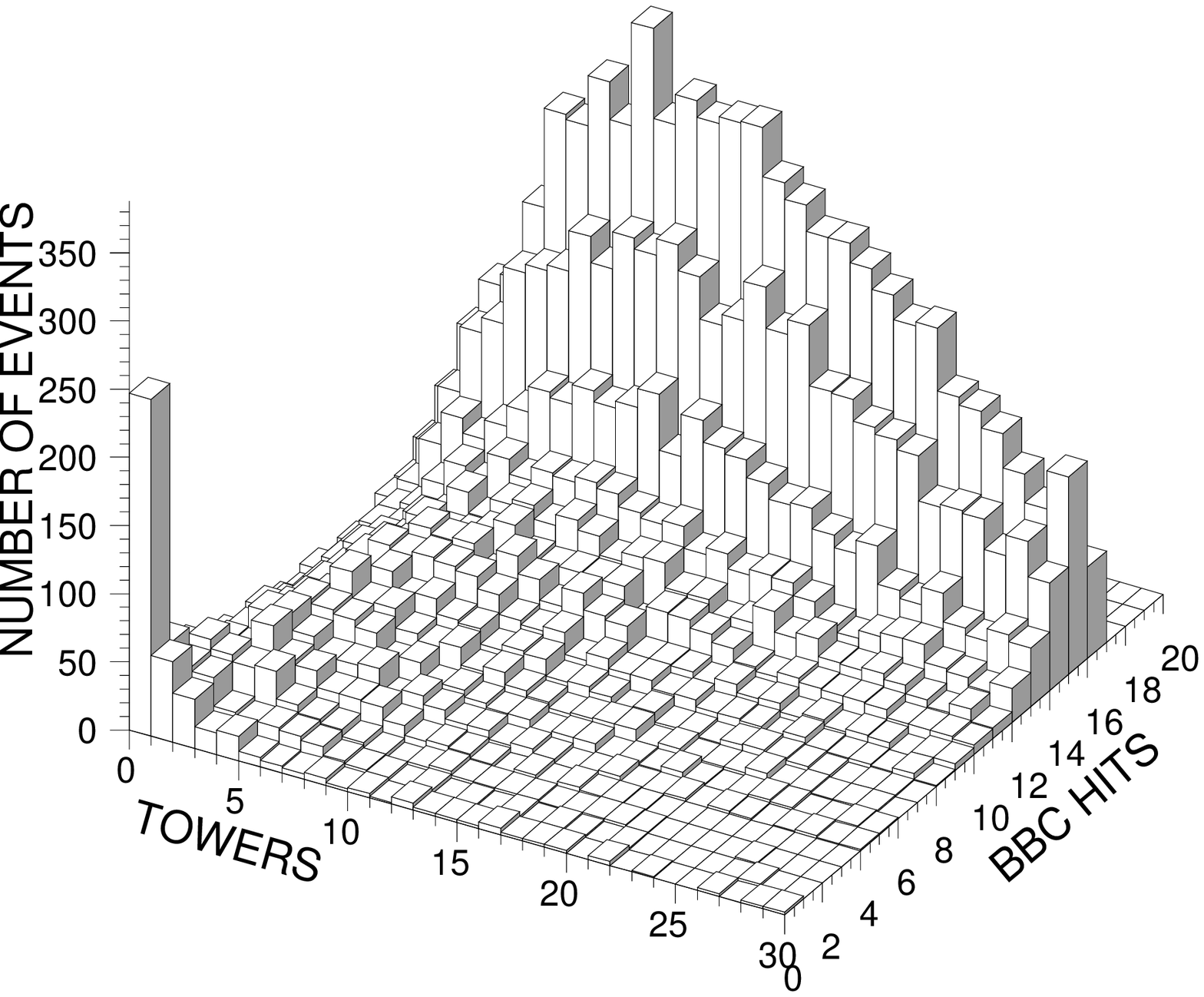,width=.45\textwidth}}
}
  \caption[]{(a) Multiplicity opposite the dijet system measured in the D\O~
electromagnetic calorimeter compared to negative binomial fits used to
estimate the non-diffractive contribution. (b) Forward calorimater tower
multiplicity versus number of hits in the BBC scintillator counter; the
diffractive peak corresponds to no hits in either detector.}
\label{tevdijet}
\end{figure}

Diffractive dijet production
has been studied at the Tevatron. 
The presence of a diffracted proton may be
identified either by a large rapidity gap on either the proton or antiproton
side or 
by directly detecting the leading proton (CDF roman pots).
In the rapidity gap analyses, 
D\O~measure uncorrected dijet rates 
with 
$E_T^{jet} > 12$~GeV and $|\eta_{jet}| > 1.6$ in coincidence with 
the multiplicity in the
electromagnetic calorimeter ($2.0 < |\eta| < 4.1$) opposite the dijet system,
as shown in Fig.~\ref{tevdijet}(a).
Similarly, CDF measure the multiplicity in the forward part
of the calorimeter ($2.4 < |\eta| < 4.2$) in coincidence with the number of
hits in the BBC scintillator counter close to the beampipe 
($3.2 < |\eta| < 5.9$), as shown in Fig.~\ref{tevdijet}(b). 
Here the jets are measured for 
$E_T^{jet} > 20$~GeV and $1.8 < |\eta_{jet}| < 3.5$ 
the diffractive events concentrate in the region 
$0.005 < \xi < 0.015$.
The ratio of diffractive dijet events
is measured by CDF to be $R_{GJJ} = 0.75 \pm 0.05 \pm 0.09$\% and 
by D\O~to be $R_{GJJ} = 0.67 \pm 0.05$\%. These preliminary figures are 
therefore in good agreement and can be used to constrain the gluon 
content of the pomeron. 
In addition, $R_{\bar{p}JJ} = 0.109 \pm 0.003 \pm 0.016$\% 
has been measured using the CDF roman pots in region of large $\xi$ 
($0.05 < \xi < 0.1$, $E_T^{jet} > 10$~GeV and $|t| < 1$~GeV$^2$).
This is a region where reggeon (quark-like) contributions are 
presumably important.

CDF have also tagged diffractive $W$ production using high-$p_T$
electrons/positrons and missing $p_T$ to tag the $W$ and then searching 
for a rapidity gap on the opposite side as in the dijet gap
analysis~\cite{cdfw}. 
The corrected ratio for diffractive/non-diffractive $W$-production is 
measured to be
$$R_W = 1.15 \pm 0.51 \pm 0.20 \%.$$

An important question when relating the various diffractive measurements  
which involve a hard scale is whether Regge factorisation, in terms of 
a pomeron flux and parton densities within the pomeron, 
is applicable. In particular, if this approach is to be useful, a
$universal$ flux is required (or a QCD description of the non-universal
correction to the flux).
In this context, the rates for diffractive processes at HERA are compared 
to those observed at the Tevatron below.

\begin{figure}[htb]
\epsfxsize=10.cm
\centering
\leavevmode
\epsfbox{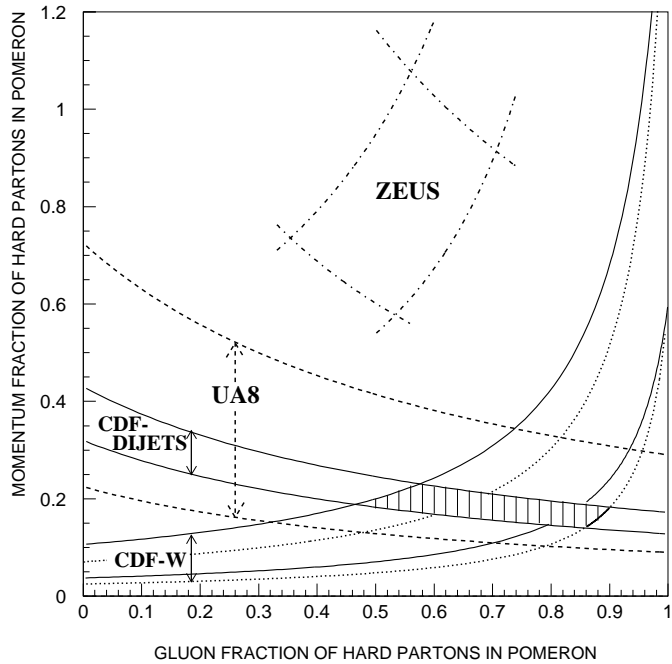}
\caption{
Momentum fraction of hard partons in the pomeron assuming a
Donnachie-Landshoff flux versus the momentum fraction carried
by the gluons in the pomeron. The ZEUS band corresponds to the allowed
region using the fits to the DIS and jet data (statistical errors only). 
The shaded band corresponds to the allowed region from the CDF dijet and $W$
analysis. The constraints given by the earlier UA8 jet data are also 
indicated.}
\label{cdfglu}
\end{figure}

We have two sets of CDF data probing the 
pomeron structure at similar momentum scales, 
$E_T^{jet}$ and $M_W$. 
Each probes the large $z$ structure of the pomeron with 
the dijet and $W$ data 
predominantly sampling the (hard) gluon and quark distributions, respectively.
In addition, we have the corresponding DIS~\cite{Zd1} and jet HERA data 
sampling the (hard) gluon and quark distributions, respectively.
In Fig.~\ref{cdfglu} the momentum 
fraction carried by the (hard) gluon, $c_g$, is plotted versus
the momentum fraction of partons in the pomeron assuming 
a Donnachie-Landshoff flux.
The CDF data are consistent with a momentum fraction carried by the 
gluons of $c_g = 0.7\pm 0.2$, in agreement with the ZEUS measurements 
of $c_g \sim 0.55 \pm 0.25$, taking into account the systematic uncertainties
due (mainly) to the estimation of the non-diffractive background.
This in turn can be compared with the H1 NLO parton distributions 
(see Fig.~\ref{h1pdfs}) which indicate $c_g \simeq 0.8$ in the high $Q^2$
region. There is therefore reasonable agreement on the parton content of the
pomeron.
However, the overall diffractive rates are significantly higher at HERA
compared to the Tevatron. This 
is reflected in the difference in the overall level of the 
ZEUS and CDF data in Fig.~\ref{cdfglu}.
This corresponds to a significantly different flux of 
pomerons (i.e. breaking of Regge factorisation) which has been predicted
in terms of QCD~\cite{soper,collins} to reduce the diffractive cross-sections
for processes which have two strongly-interacting initial state hadrons. 
These effects are not apparent in the HERA data, where a virtual photon
or a (predominantly) direct photon participate in the hard scattering
process.

Double pomeron exchange, where the $p$ and $\bar{p}$ remain intact, has also 
been studied by CDF and D\O~for their dijet samples. This process should be
directly sensitive to Regge-factorisation breaking effects. Both experiments
find a ratio of hard double pomeron exchange events to non-diffractive events
of $\simeq 10^{-6}$. This is consistent with independent dissociation of the
$p$ and $\bar{p}$, with probabilities $\simeq 10^{-3}$, but further studies
are required to establish these rates and determine whether these can be
explained by the factorisation-breaking calculations~\cite{collins}. 

\section{Conclusions}

The soft pomeron does not describe $all$ diffractive data measured at HERA.
As the photon virtuality or the vector meson mass increase 
a new dependence on $W^2$ emerges.
As we investigate the pomeron more closely, a new type of dynamical pomeron
may begin to play a r\^ole; a dynamical pomeron whose structure is being 
measured in DIS. 
These data are consistent with a partonic description of the exchanged object
which may be described by pQCD.

The cross-sections for hard diffractive processes at HERA can be compared 
to those observed at the Tevatron.
All data are consistent with a significant gluon contribution of the partons 
within the pomeron i.e. QCD factorisation appears to be observed.
However, a universal pomeron flux does not describe the observed rates
i.e. Regge factorisation does not apply.
This is qualitatively expected from QCD corrections to the rates, but the 
effects are large and non-perturbative.

The experimental work focuses on extending
the lever arms and increasing the precision in $t$, $M^2$, $W^2$ and $Q^2$
in order to explore this new structure. Before more precise tests 
can be made, further theoretical and experimental
input is required to reduce the uncertainties due to 
non-diffractive backgrounds and proton dissociation
as well as the treatment of $F_L$ and radiative corrections.

\section*{Acknowledgements}                                          
                                                                                
The results presented in this talk are a summary of significant 
developments at HERA and the Tevatron in the study of diffraction.
The financial support of the DESY Directorate and PPARC allowed me to 
participate in this research, whilst based at DESY, for which I am very
grateful.
Many thanks to Halina Abramowicz, Nick Brook, Allen Caldwell, Nicolo Cartiglia,
Malcolm Derrick, John Dainton, Martin Erdmann, Elisabetta Gallo, Peppe
Iacabucci, Michael Kuhlen, Aharon Levy, Jason McFall, Paul Newman, Juan Puga, 
David Saxon, Laurel Sinclair, Ian Skillicorn, Juan Terron and Robert Waugh 
for their encouragement, enthusiasm, help and advice.
It is a pleasure to thank the organisers for their additional 
financial support and a very enjoyable summer school. 
\vspace{2.0cm}

\end{document}